\documentclass[prb,amsmath,amssymb,twocolumn]{revtex4-1}
\usepackage[normalem]{ulem}
\usepackage{amsmath,amssymb}
\usepackage[bookmarks=true,colorlinks,linkcolor=blue,urlcolor=blue,citecolor=blue]{hyperref}
\usepackage{graphicx,amsmath,amsfonts,dsfont}
\usepackage[usenames]{color}
\usepackage{epstopdf}
\usepackage{verbatim}
\usepackage{amsthm}
\usepackage{enumitem}
\hypersetup{colorlinks,linkcolor=blue,filecolor=green,urlcolor=blue,citecolor=blue}
\usepackage[titletoc,toc]{appendix}

\newcommand{\dtt}{\tilde{\partial}_t}

\newcommand{\dip}{\mathrm{det}}

\newcommand{\ezr}{\eta_{\mathit{ZR}}}
\newcommand{\ezi}{\eta_{\mathit{ZI}}}

\renewcommand{\Re}{\mathop{\mathrm{Re}}}
\renewcommand{\Im}{\mathop{\mathrm{Im}}}

\DeclareMathOperator{\tr}{tr}

\newcommand{\be}{\begin{equation}}
\newcommand{\ee}{\end{equation}}

\newcommand{\syse}{\end{array}\right.}

\newcommand{\ag}[1]{}

\begin{document}

\title{Quantum dynamical  field theory \\
for non-equilibrium phase transitions in driven open systems}

\author{Jamir Marino$^{1,2}$}
\author{Sebastian Diehl$^{1,2}$}

\affiliation{$^1$Institute of Theoretical Physics, TU Dresden, D-01062 Dresden, Germany\\
$^2$Institute of Theoretical Physics, University of Cologne, D-50937 Cologne, Germany}

\begin{abstract}

We develop a  quantum dynamical field theory for studying  phase transitions in  driven open systems coupled to Markovian noise, where non-linear noise effects and  fluctuations beyond semiclassical approximations influence the critical behaviour. 
We systematically compare the diagrammatics, the properties of the renormalization group flow and the structure of the fixed points, of the novel quantum dynamical field theory and of its semi-classical counterpart, which is employed to characterise  dynamical criticality in three dimensional driven-dissipative condensates. 
As an application, we perform the Keldysh Functional Renormalization of a one dimensional driven open Bose gas, where a tailored diffusion Markov noise realises an analog of quantum criticality for driven-dissipative condensation. We find that the associated non-equilibrium quantum phase transition does not map into the critical behaviour of its three dimensional classical driven counterpart.

\end{abstract}


\date{\today}

\maketitle

\section{Introduction}

Universality classes constitute an important concept in the modern understanding of statistical mechanics and condensed matter physics.  The divergence of the correlation length at a phase transition  indicates that specific microscopic details are inessential in order to characterize the physical properties of the system, and few independent critical exponents encode all the relevant physical information at long wavelengths or time scales. Specifically, such exponents characterize the anomalous power law scaling behaviour of several observables and   the thermodynamic singularities close to the critical point, usually separating two stable phases \cite{Amit/Martin-Mayor, Zinn-Justin, Cardy}. 

A textbook example is provided by the classical Ising model, whose critical behaviour is driven by thermal fluctuations.
Insensitivity to microscopic details  is manifest in the exact coincidence of the critical exponents of spin magnetic systems that share the same symmetry of the Ising Hamiltonian and have the same lattice dimensionality. This  circumstance has made the classification of all possible universality classes a key challenge in modern physics, stimulating a field theoretical formulation of phase transitions, in the framework of renormalization group theory \cite{Amit/Martin-Mayor, Zinn-Justin, Cardy}.

The renormalization group provides a powerful tool to extract not only static, but also dynamic properties near a critical point. Extending  scaling behaviour to time dependent correlation functions, and introducing the dynamical critical exponent, $z$, it is possible to extract the universal large distance and long time scaling behaviour of correlation and response functions, enlarging the concept of universality to equilibrium critical dynamics \cite{tauberbook, HHRev, enz79:_field}. 

The dynamics of classical equilibrium phase transitions is sensitive to the presence or absence of conservation laws: models with different conserved quantities display a different dynamical critical exponent, despite them obeying the same equilibrium statistical mechanics, and  sharing the same static critical exponents. The  classification of  the dynamics of phase transitions pioneered by Hohenberg and Halperin \cite{hohenberg77:_theor}, is based on the Martin-Siggia-Rose-Janssen-deDominicis (MSRJD) method \cite{martin73:_statis_dynam_class_system, enz79:_field, Janssen1976, DeDom76, DeDom78, tauberbook, kamenevbook}, which translates  a stochastic differential equation into  a field theory formulated in terms of a path integral. The Martin-Siggia-Rose action is tailored to study classical phase transitions where quantum fluctuations can be neglected, and where the noise is Gaussian and short ranged in time and space.

Prototypical examples of non-equilibrium classical critical systems range from reaction-diffusion models \cite{Doi1976,Peliti1985,Cardy1996} (e.g. directed percolation \cite{Obukhov1980,Hinrichsen2000})  and driven-diffusive lattice gases \cite{Zia1995}, to  random surface growth dynamics, like Kardar-Parisi-Zhang (KPZ) models ~\cite{Kardar1986, krug91, Stanley95, halpinhealy95, Krug97, Lassig98}, encompassing  aging dynamics in relaxational models~\cite{Janssen1989, janssen92, Calabrese2005, henkel07:_field_theor_approac_noneq_dynam}.

Complementary to equilibrium classical phase transitions, a quantum many body system can encounter a phase transition at zero temperature, when a parameter of  its Hamiltonian is  changed, in full analogy to temperature in classical critical phenomena \cite{sondhi, Sachdev, vojta}. In such phase transitions dynamic and static critical properties are intertwined, with time playing the role of an extra dimension. As a remarkable consequence,  $d$ dimensional quantum phase transitions fall in the universality class of their $d+z$ finite temperature classical counterparts. 

Thermal fluctuations are detrimental for quantum critical behaviour, which survives asymptotically only at zero temperature -- in other words, temperature is a relevant perturbation at a quantum critical point. However, 
quantum criticality can  still be relevant at  finite $T$, for a frequency regime, $\hbar\omega\gg k_BT$, above the quantum critical point.

The current state of the art \cite{Sieberer2015, Tauber16} on non-equilibrium classical phase transitions  and on  quantum critical phenomena, poses several intriguing, yet natural, questions on the existence and nature of a non-equilibrium analog of quantum criticality: Which are the universality classes of non-equilibrium critical quantum many body systems? Do these universality classes possesses non-equilibrium classical  counterparts? Which are the relevant physical perturbations capable of destroying non-equilibrium quantum critical behaviour, as temperature does at equilibrium? Which is the effective description in terms of a non-equilibrium quantum field theory of these novel critical points? Can these novel critical phenomena be understood in terms of a renormalization group approach? 

The search for non-equilibrium universality in quantum systems has been triggered by an influential work on the study of critical behaviour in
itinerant electron magnets coupled to two external reservoirs, and   driven  by a non-equilibrium steady current~\cite{Mitra2006}. 
Later,  the emergence of  non-thermal fixed points  has been documented in a series of works on  the study of superfluid turbulence \cite{Berges2008, Scheppach2010, Nowak2011, Nowak2012, Schole2012, Berges2012, Orioli2015}. 

Concerning the proper onset of non-equilibrium quantum criticality, first studies focused on systems driven by non-Markovian noise~\cite{DallaTorre2010, DallaTorreDemler2012}, while, more recently, instances of universality have been searched in the aging dynamics of isolated~\cite{Chiocchetta2015, Maraga2015, Chiocchetta2016} or open~\cite{Gagel2014,Gagel2015,Buchhold2014} quantum systems (see also  Refs. \onlinecite{Eckstein2009,Schiro2010, Sciolla2010, Sciolla2013, Smacchia2015} for other examples of dynamical critical behaviour after a quench).
 
In addition to these relevant examples, driven-dissipative quantum many body systems, i.e. many-particle systems where quantum coherent dynamics and dissipative effects occur on the same footing, can constitute a promising platform~\cite{hartmann08, ates12, olmos13, lee13, sarkar13, janot13, Sadri2014, Elliott2015} to initiate a systematic classification of quantum dynamical criticality~\cite{DallaTorreDiehl2013,Sieberer2013, Tony2014, Garr2015, Altman2015, Nagy2015, Koch2015, Marino2016, Dunnett2016}. 

Experiments and theory ranging from polariton condensates \cite{carusotto05, Yamamoto10, Byrnes14, keeling04, Kasp2006, keeling10}, over cold atoms in optical cavities or driven-dissipative microcavity arrays \cite{Hood2000, esslingerdicke, esslingerdicke3, Nagy2010, Nagy10, Szirmai2010, clarke-nature-453-1031, Mottl2011, Maschler2011, houck12, koch13, Mabuchi2013, esslingerQEDreview, Nagy2015}, to trapped ions \cite{blatt12, britton12},  have triggered  the interest in driven-dissipative systems, where  quantum dynamics and  dissipation can  interplay, and open the door to novel phase of matter, proper of a non-equilibrium setting. 

In this context, several theoretical studies\cite{Mitra2006, Diehl08, diehl10:_dynam_phase_trans_instab_open, dalla10:_quant, torre13:_keldy, Oztop2012, wouters07} (three-dimensional driven-dissipative condensates, open electronic systems, noisy one-dimensional quantum liquids, driven-open non-linear Dicke models) have reported the emergence of an effective infrared thermal behaviour and generic loss of quantum coherence close to criticality. Such circumstance does not contradict the microscopic non-equilibrium nature of these systems:  breaking of detailed balance still occurs in the ultraviolet (microscopic) features of driven-dissipative 
 steady states. 

In this work, we show that both non-equilibrium and quantum features can be simultaneously present and persist in the low frequency properties of a critical one-dimensional driven open Bose gas, and no effective equilibrium description is applicable for this novel type of non-equilibrium quantum criticality. Such novel critical regime can be achieved tailoring a quantum diffusion term described microscopically by a Lindblad master equation; indeed, this novel diffusion Markov noise mimics the canonical scaling close to the critical point of a zero temperature open bosonic  system. However,  for this novel critical behaviour, non-linear quantum effects beyond semiclassical approximations and non-Gaussian noise terms are relevant in a renormalization group sense. This requires  the development of a  quantum dynamical field theory beyond the Martin-Siggia-Rose-Janssen-deDominicis (MSRJD) approach \cite{tauberbook, kamenevbook}. Such construction extends to the non-equilibrium quantum domain the notion of classical dynamical field theory pioneered by Hohenberg and Halperin \cite{HHRev} to study dynamical critical phenomena, and it constitutes the core of the present work. 

\subsection{Summary of the results}

This paper constitutes a follow-up of our work in Ref. \onlinecite{Marino2016}. We now summarise the main achievements of Ref. \onlinecite{Marino2016} and of the current paper, highlighting the inedited results.\\

\emph{Quantum dynamical field theory --} In Sec. \ref{FRGSec} we elaborate  on the field theory developed to extract the critical behaviour of the driven-dissipative Bose gas in the quantum scaling regime. We coin the term \emph{ quantum dynamical field theory} in opposition to the notion of \emph{classical dynamical field theory}, usually employed to inspect classical non-equilibrium critical behaviour, or even the simpler case of equilibrium critical dynamics. For instance, the entire Hohenberg-Halperin classification\cite{HHRev} of critical dynamics on the basis of conservation laws, is grounded on  classical dynamical field theories formulated in terms of the  Martin-Siggia-Rose-Janssen-deDominicis formalism \cite{kamenevbook, tauberbook}. The MSRJD functional is an equivalent functional integral representation of stochastic Langevin dynamics, where the coarse-grained order parameter field associated to the phase transition is accompanied by a noise or response field.


In non-equilibrium (KPZ models, directed percolation, etc) and equilibrium  (e.g. purely relaxation models coupled to a bath) classical dynamical field theories, the relevant terms in a renormalization group sense are  the ones up to linear order in the noise field for the deterministic part of the action, and quadratic in the noise field \cite{kamenevbook, tauberbook, Altland/Simons}. The diffusion Markov noise tailored for the driven open Bose gas, instead opens the door to a scaling regime where quartic terms compatible with the symmetries of the model and involving all the possible non-linear occurrences of the response field,  have to be included in a renormalization group formulation, in order to properly capture the critical regime and the associated infrared fixed point. These quartic operators represent mathematically the inclusion of  fluctuations, beyond previously adopted semi-classical expansions for driven-dissipative criticality\cite{Sieberer2014}.

We are interested in finding an interacting, or Wilson-Fisher\cite{Amit/Martin-Mayor}, fixed point (FP) of the renormalization group (RG) flow, i.e. a fixed point of the RG beta functions with non-vanishing values of the couplings constants. An interacting fixed point indicates non-trivial scaling behaviour at criticality which cannot be predicted from the  knowledge of the Gaussian theory (mean-field analysis) \cite{Amit/Martin-Mayor, Zinn-Justin, Cardy}. In particular, for the interacting fixed point critical exponents are usually irrational  numbers, while for the latter case they are rational. In Sec. \ref{FRGWettSec} we report on the Functional Renormalization \cite{wetterich08:_funct} of the action, $S_Q$, associated to the quantum dynamical field theory, and employed to find the interacting fixed point. We briefly summarise in two Appendices  some technical aspects of interest for experts on functional renormalization group (FRG) methods.\\

\emph{Driven-dissipative quantum Bose condensation--} The entire paper is built on a comparison between the critical regime induced by fine tuning the spectral gap and the momentum dependent Markov noise level in the  driven open Bose gas in one dimension, and its three dimensional counterpart, which is instead coupled to flat Markovian noise. This comparison is grounded on the fact that the former realises a driven analog of a zero temperature quantum scaling regime, while the latter is analogous to a finite temperature situation in terms of the canonical scaling. The critical point separates in both cases an ordered BEC-condensed phase from a disordered one. We will sometimes refer to criticality in the former as the driven-dissipative quantum condensation regime, or \emph{driven Markovian quantum criticality}, while we mention the latter as the driven-dissipative semi-classical condensation regime, or \emph{driven Markovian semi-classical criticality} (developed in Refs. \cite{Sieberer2013, Sieberer2014}). The driven-dissipative bosonic platform used to implement such out-of-equilibrium BEC condensation regimes, the properties and implementation of the Markov diffusion noise, are discussed in Section \ref{DQFT}.

It should be stressed that both situations constitute a non-equilibrium setting where both  energy and total number of particles are not conserved, since the system is driven out-of-equilibrium by continuous pump and loss of particles from external reservoirs. This should be contrasted and seen as a non-equilibrium extension of the paradigmatic case of criticality in open (number non-conserving) systems at equilibrium with an ohmic thermal bath of quantum harmonic oscillators, the so-called Caldeira-Leggett model \cite{caldeira, kamenevbook}. In that case,  the bath affects, in general, with  non-Markovian noise level the equilibrium critical dynamics of the open system.
Only in the high temperature phase the Caldeira-Leggett model becomes coupled to \emph{Markovian} thermal fluctuations, while at zero-temperature the bath contributes with \emph{non-Markovian} noise \cite{kamenevbook}. The comparison between criticality in a Caldeira-Leggett model and in  driven-dissipative Bose gases is elucidated in Sec. \ref{QCanonical}.\\

\emph{Absence of decoherence at long-wavelengths --} Besides detailing the findings of Ref. \onlinecite{Marino2016}, we compare  more explicitly  the fixed point found in Ref. \onlinecite{Marino2016}, where coherent and dissipative effects compete, and the purely dissipative fixed point (where quantum coherent effects have completely washed out), which describes the regime of semi-classical driven-dissipative condensation of Refs. \onlinecite{Sieberer2013, Sieberer2014}. In particular, the latter is never realised in the quantum scaling regime induced by diffusion Markov noise, as shown in Sec. \ref{decohsec}: qualitatively, a certain amount of quantum coherence is expected to persist at the fixed point, if an analog of zero-temperature quantum criticality is to be  established (see for instance Ref. \onlinecite{sondhi}). The consequences of absence of decoherence are evident also on the physical properties of the system at the quantum critical point: the critical dispersion relation is not  dominated by leading diffusive modes (see Sec. \ref{modesec}), rather coherent propagation and dissipative motion concur on the same footing, i.e. they are characterised by the same anomalous dimensions. This  results from a degeneracy of anomalous exponents controlling the scaling of coherent and dissipative kinetic coefficients. As a further signature of absence of decoherence at criticality,    the spectral density exhibits RG limit-cycle oscillations, as we discuss in Sec. \ref{decohsec}.\\

\emph{Non-thermal character of the infrared distribution  function --} We corroborated the results of Ref. \onlinecite{Marino2016} with a formal discussion, in  Sec. \ref{constrsec}, on the properties that RG beta functions should possess in order to lead the RG flow towards an effectively thermal FP. From this analysis, we provide further evidence that the novel fixed point does not fall into any extension of know  equilibrium universality classes (Sec. \ref{FDRsec}). This is an important difference with the fixed point of semi-classical driven-dissipative condensation, where an effective thermal character of the distribution function, emerges  at infrared scales  \cite{Mitra2006, Diehl08, diehl10:_dynam_phase_trans_instab_open, dalla10:_quant, torre13:_keldy, Oztop2012, wouters07, Sieberer2013, Sieberer2014}. It should be reminded that the eventual presence of effective thermalization at infrared scales  is not in contradiction  with the out-of-equilibrium nature of the system, since the high-momentum tails of the distribution function still carry signatures of its driven condition \cite{torre13:_keldy}. Moreover, even if correlation functions display thermal character in the vicinity of effectively thermal fixed points of driven open systems,  a new independent critical exponent in the response functions reflects the presence or absence of energy conservation, and  distinguishes equilibrium from non-equilibrium critical behaviour \cite{Sieberer2013}. \\

\emph{Quantum and semi-classical driven Markovian criticality are not in the same universality class --} Extending the results of Ref. \onlinecite{Marino2016}, we provide in Sec. \ref{absmap} a diagrammatic argument to corroborate the result that the universality classes of the quantum and semi-classical critical condensation regimes do not coincide. This difference is  grounded on the lack of restoration of a thermal fluctuation-dissipation relation at infrared scales,  which can be formulated as a symmetry of the Keldysh action.  The lack of this symmetry in the  effective action of driven Markovian quantum criticality, is a straightforward reason for the mismatch between the two universality classes. \\

 \emph{Extension of the quantum scaling regime--} We finally estimate the momentum window where the novel critical exponents associated to the quantum fixed point can be observed (for instance, in the scaling behaviour of correlation functions). A flat homogeneous Markovian noise level can never be completely eradicated in a driven-dissipative gas, and at low momenta it will ultimately overcome the diffusion Markov noise, since this is gapless and quadratic in momentum. An upper bound for the momentum scale  where this occurs, and accordingly the system leaves the quantum scaling regime, is provided in Sec. \ref{secestim} . We refer in the following to this momentum scale as Markov scale, $\Lambda_M$. While $\Lambda_M$ is  analogous to the de-Broglie length, $\Lambda_{dB}\sim1/T^{1/z}$, for finite temperature quantum phase transitions, the estimate of $\Lambda_M$ requires the knowledge of the infrared action, since the Markov noise level flows under RG, in contrast to temperature in equilibrium systems: this allows, at least in principle, to set the de-Broglie length from the outset in a finite temperature quantum phase transitions \cite{sondhi, Sachdev, vojta}, and fine tune the extent of the equilibrium quantum scaling regime. The extension of the quantum scaling regime is obtained by computing a second momentum scale, known as Ginzburg scale \cite{Amit/Martin-Mayor}, $\Lambda_G$, below which the scaling behaviour dictated by the novel interacting fixed point can be observed, since for momenta close to $\Lambda_G$ and below, mean-field theory breaks down and  trivial Gaussian scaling does not hold anymore. It is important that the two scales, $\Lambda_G$ and $\Lambda_M$, are separated in order to resolve in an experiment the novel scaling predicted by our RG analysis. The estimate of $\Lambda_G$ is  discussed in Sec. \ref{secestimG}, and it requires a standard perturbative computation \cite{Amit/Martin-Mayor}.

\section{Driven dissipative Bose gas 
and diffusion Markov noise}
\label{sec:DD}

In this first Section we introduce our model, a one dimensional driven open Bose gas in the presence of a diffusion Markov noise.
We first write the model in terms of a second quantised master equation, and  we then convert it into a Keldysh functional integral, which is technically more suitable in order to search for long-wavelength universal properties at the critical point.
In this Keldysh field theory framework, we discuss in detail how the presence of a diffusion Markov noise in the driven open Bose gas, can induce a regime of quantum condensation out-of-equilibrium, which constitutes a non-equilibrium analog of bosonic zero-temperature quantum criticality for driven systems. 
We end the Section with an implementation of this diffusion noise with a quantum optics architecture of superconducting qubits coupled to microwave resonators.
\subsection{Physics of the Lindblad equation}
\label{microscop}

The open quantum dynamics of the gas (in a generic number of spatial dimensions, $d$) is described  by the Markovian quantum master equation for the density matrix, $\hat{\rho}$, 

\begin{equation}\label{master}
\partial_t\hat{\rho}=-i[H,\hat{\rho}]+\mathcal{L}[\hat{\rho}].
\end{equation}

Coherent dynamics is encoded in the Hamiltonian, $H$, and dissipative processes enter  the Lindbladian super-operator, $\mathcal{L}$. 
We assume in the following that the environment is  Markovian  and  choose coherent and dissipative processes according to typical  set-ups for polariton or circuit QED experiments.
Crucial physical ingredients in our analysis are the driven and open nature of the system under study: They dictate the absence of both energy and particle conservation. While  the latter is compatible with equilibrium conditions, the absence of energy conservation causes non-equilibrium conditions even in the stationary state \cite{Sieberer2014, Sieberer2015} (for a driven Bose dynamics preserving the particle number, see for instance Ref. \onlinecite{Buchhold2014}).

We now detail the explicit form of the master equation \eqref{master}.
We represent the bosonic degrees of freedom of the gas in second quantization, through the bosonic  field annihilation and  creation operators, $\hat{\phi}(x)$, and, $\hat{\phi}(x)^\dag$. With such a choice, contact interactions of strength $\lambda$ among bosons are written in the Hamiltonian, as
\begin{equation}\label{hamilt}
H=\int_x\hat{\phi}^\dag(x)\left(-\frac{\nabla^2}{2m}\right)\hat{\phi}(x)+\frac{\lambda}{2}\int_{x}\hat{\phi}^\dag(x)^2\hat{\phi}(x)^2,
\end{equation}
where we did not include the chemical potential of the Bose gas, since the density of the system is kept at a constant value by the balance of particle loss and gain. 

This aspect can be elucidated providing details on the explicit form  of  the Liouvillian, which is given by the sum of four dissipative channels,

\begin{equation}\label{lindbladmaster}
\mathcal{L}=\sum_{a=p,l,t,d} \mathcal{L}_a.
\end{equation}
%
%
In particular, each $\mathcal{L}_a$ has Lindblad form
\begin{equation}
\mathcal{L}_a[\rho]=\gamma_{a}\int_{x}\left(\hat{L}_a(x)\hat{\rho} {\hat{L}^\dag_a}(x)-\frac{1}{2}\left \{\hat{L}^\dag_a(x)\hat{L}_a(x),\hat{\rho}\right\}\right).
\end{equation}
The Lindblad operators 
\be
\hat{L}_{p}(x)=\hat{\phi}^\dag(x), \quad \hat{L}_{l}(x)=\hat{\phi}(x),
\ee
incoherently create and destroy single particles in a given space point, $x$,  with rates $\gamma_p$ and $\gamma_l$, and the Lindblad  operator 
 \be
 \hat{L}_{t}(x)=\hat{\phi}(x)^2
 \ee
is responsible for the local incoherent simultaneous destruction of two particles with rate $\gamma_t$. 
The local jump operators $ \hat{L}_{a}(x)$ ($a=\{p, l, t\}$) dictate the driven, non-energy conserving, nature of the system, and they are also responsible for the lack of  particle number conservation.
For the sake of clarity, we notice that in this Subsection the label $x$ denotes a generic $d$-dimensional set of spatial coordinates, even if for a wide part of the work we will consider one-dimensional Bose gases. The  dissipative channels discussed up to now constitute the main ingredients to achieve Bose condensation in a three dimensional driven-dissipative Bose gas.
  
The last Lindblad operator of the Liouvillian \eqref{lindbladmaster}
  \be\label{diffusslind}
  \hat{L}_{d}(x)=\partial_x\hat{\phi}(x),
  \ee
describes single particle diffusion with rate $\gamma_d$, and it results, indeed, crucial in order to achieve an out-of-equilibrium analog of quantum criticality in the one-dimensional driven-dissipative Bose gas. 
We discuss in full detail the role of this operator  later in this Section, while we now focus on a  mean-field analysis of   driven-dissipative Bose condensation.

\subsection{Driven-dissipative condensation}

\begin{figure}[t!]\centering
\includegraphics[width=5.5cm]
{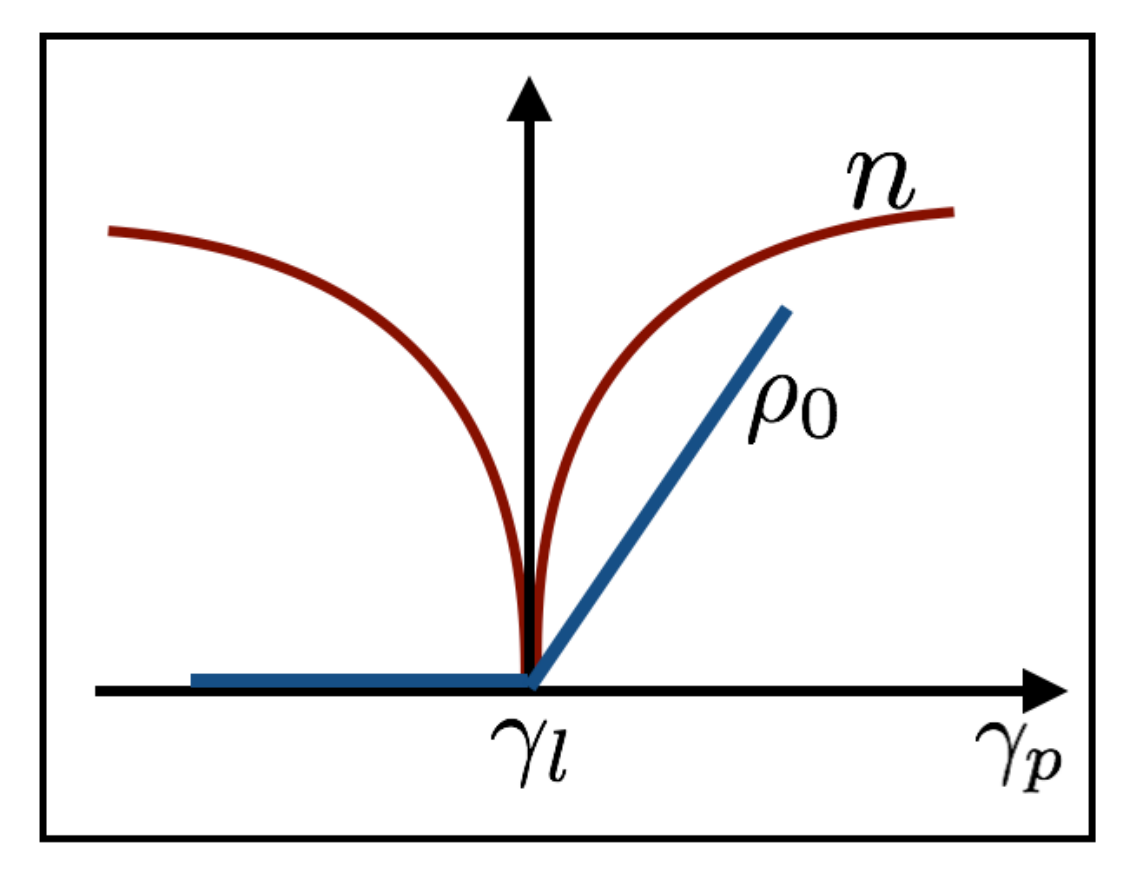}
 \caption{(Color online) Condensate density, $\rho_0\propto-\chi$, and number particle density, $n\propto\sqrt{|\chi|}$, in the vicinity of the driven-dissipative transition. 
}
\label{transfig}
\end{figure}

\subsubsection{Three dimensions}

We start reviewing the mean-field phase diagram of  driven-dissipative Bose condensation in three dimensions (see for a more exhaustive analysis Ref. \onlinecite{Sieberer2014}).

When the single particle pump exceeds the single particle loss, $\gamma_p>\gamma_l$, the Bose gas  settles in a stationary state, $\hat{\rho}_{ss}$, in which a finite Bose-Einstein condensate amplitude  builds up,
\be
\langle\hat{\phi}(x) \rangle_{ss}=\operatorname{Tr}[\hat{\rho}_{ss}\hat{\phi}(x)]=\phi_0\neq0.
\ee
In such condensed phase the $U(1)$ symmetry of Eq. \eqref{master}, described by the $U(1)$ rotation of field operators (with a given angle $\alpha$), $\hat{\phi}(x)\to\hat{\phi}(x) e^{i\alpha} $, is broken. 

The condensate is stabilised by the presence of two-body losses, as it can be seen in a mean-field approximation, based on a coherent state ansatz for $\hat{\rho}$ in the master equation \eqref{master}; such approximation  results \cite{Sieberer2014} in a  noiseless evolution for the homogeneous condensate amplitude, $\phi$,
\begin{equation}\label{GPE}
i\partial_t \phi=[\lambda|\phi|^2-\frac{i}{2}(\chi+2\gamma_t|\phi|^2)]\phi,
\end{equation}
where $\chi=(\gamma_l-\gamma_p)/2$. This quantity measures the distance from the phase transition (occurring at $\chi=0$, or $\gamma_p\to\gamma_l$) and therefore we refer to it as a spectral mass or gap.
For $\gamma_p>\gamma_l$, we can search for a solution (see for more details Ref. \onlinecite{Sieberer2014}) of the form $\phi=\phi_0e^{-i\mu t}$, where the condensate density is found setting to zero the imaginary part of the right-hand side of Eq. \eqref{GPE}, choosing $\mu=\lambda|\phi_0|^2$,
\be\label{densitymf}
\rho\equiv|\phi_0|^2=\frac{\gamma_p-\gamma_l}{2\gamma_t}.
\ee
In this case, the gap is negative, $\chi<0$, and the condensate density can be traded one for the other, since $\chi\sim-\rho_0\gamma_t$

The steady-state phase diagram of a three dimensional driven-dissipative Bose gas is therefore characterised by an ordered phase with a spontaneous symmetry breaking of the global $U(1)$ symmetry, and by a disordered, symmetric one. 
The two phases are distinguished by tuning the single-particle pump rate, $\gamma_p$, above or below, $\gamma_l$, the one-body loss rate.
%
%

The impact of critical fluctuations beyond the mean-field picture, has been already explored (see Ref. \onlinecite{Sieberer13, Sieberer2014}) for a three dimensional driven-dissipative Bose gas in the presence of a Markovian environmental noise, $\xi$, delta correlated in space
\be\label{noisecorrflat}
\langle \xi^*(t,x)\xi(t,x') \rangle=\gamma\delta(t-t')\delta(x-x'),
\ee
where the variance, $\gamma$, depends on the sum of single body loss and pump rates, $\gamma=\gamma_l+\gamma_p$.

The noise occurs in a generalization of Eq. \eqref{GPE} for the evolution of the space-dependent condensate amplitude, $\phi(x)$, which is a stochastic Gross-Pitaevskii equation   

\be\label{evolstoc}
i\partial_t \phi=[-(K_R-iK_I)\Delta-\mu-i\chi +2(\lambda-i\gamma_t)|\phi|^2]\phi+\xi,
\ee 
already widely employed  in literature as a model for exciton-polariton condensates \cite{carusotto05, szymaifmmode06, keeling08,keeling10, wouters07,wouters10}. In Eq. \eqref{evolstoc}, $\mu$ sets a rotating frame for the gas and $K_R\equiv1/2m$ (associated to coherent propagation) is the inverse effective mass of the bosons.
As already shown after Eq. \eqref{GPE}, $\mu$ can be gauged away.

We recall that the Langevin dynamics encoded in Eq. \eqref{evolstoc}, can be  derived from a semi-classical Keldysh path integral description~\cite{Sieberer2015} (i.e. within the MSRJD formalism). 
\\

While the condensate density scales with the spectral mass  gap as $\rho_0\propto -\chi$ in the ordered phase, and it vanishes in the symmetric one, the particle density scales as $n\propto \sqrt{|\chi|}$, and it is non-vanishing also in the symmetric phase.

The particle density can be easily computed through the Keldysh Green's function\cite{Sieberer2015} (see Eq. \eqref{green} for the definition of $G^K$, and the action discussed  in Sec. \ref{introSQ} for proper definition of the quantities involved), 
\be\begin{split}\label{densitypart}
n&=\frac{i}{2}\operatorname{Tr}_{2\times2}\int d^d\mathbf{q}\left(\int d\omega G^K(Q)-1\right)= \\
&\int d^d\mathbf{q}\left(\int d\omega  \frac{\gamma+2\gamma_dq^2}{(\omega-K_Rq^2)^2+(K_Iq^2+\chi)^2}-1\right), 
\end{split}\ee 
which is written in $d$ spatial dimensions. Setting $d=3$, and assuming $\gamma\gg2\gamma_dq^2$ close to the transition (in the spirit of a derivative expansion), we find $n\propto\frac{\gamma_p}{\gamma_d^{3/2}}\sqrt{\chi}$ for the  symmetric phase\footnote{We subtracted an ultraviolet divergence, which is interpreted as a zero point shift for each momentum mode.}, while  in the ordered phase  tree-level shifts due to the presence of a condensate have to be included\cite{Sieberer2014} in order to cure the singular behaviour of Eq. \eqref{densitypart} for $\chi<0$. Since the latter  computation is lengthier, we did  not report it here: a summary of the results is contained in Fig. \ref{transfig}. 

One of the main conclusions of this work is that dimensionality and the type of Markov noise  considered, determine respectively a regime of classical and quantum condensation out-of-equilibrium.

In the limit $\gamma_p\to\gamma_l$, with $\gamma_p$ and $\gamma_l$ finite, the critical point of  three dimensional driven-dissipative condensation is reached, and a semi-classical field theory formulation of the stochastic Gross-Pitaevskii, Eq. \eqref{evolstoc}, is sufficient to capture the universal properties of the transition, as reported previously in Refs. \onlinecite{Sieberer2013, Sieberer2014}.

\subsubsection{One dimension}

%

%
From now on, we  focus instead on  the one-dimensional limit of  driven-dissipative Bose-Einstein condensation, which is achieved in the presence of a strong diffusion Markov noise determined by  the Lindblad operator $\hat{L}_d(x)$. 
If in addition to $\gamma_p\to\gamma_l$, also the limit $2\gamma_dq^2\gg\gamma$ is considered  (i.e. $\gamma_l+\gamma_p\to0$ as well), the  Markov diffusion  prevails in the noise level, and the phase space of the one-dimensional driven open gas experiences an effective dimensional enhancement (as in zero-temperature quantum criticality), since the quadratic scaling of the  noise compensates for the lack of the three dimensional integration measure in integrals involving the noise. This mechanism opens the door to a quantum condensation regime in this low dimensional setting, as we discuss in detail  in Sec. \ref{QCanonical}.

%
Also in this one-dimensional setting the density of particles does not vanish in the symmetric phase: setting $d=1$ and $\gamma\to0$ in Eq. \eqref{densitypart}, we find again $n\propto\sqrt{|\chi|}$ (furthermore, $\rho_0\sim-\chi$, see again Fig. \ref{transfig}). This an important occurrence, since otherwise one would have a  transition from the vacuum towards a finite density phase, as it occurs for the Mott insulator to  superfluid transition at $T=0$ in the Bose-Hubbard model. Such equilibrium quantum phase transition  exhibits rational critical exponents due to particle number conservation -- away from the particle-hole symmetric points where the dynamical critical exponent is $z=2$ (see for a comprehensive discussion Ref. \onlinecite{Sachdev}). For the  phase transition studied in this work, instead, particle number is not conserved, and the critical state associated to this non-equilibrium transition is characterised by the condensate density  \eqref{densitymf} in the limit $\gamma_p\to\gamma_l$, resulting from continuous pump and loss of particles. On the basis of  these  arguments, we can  expect for these driven-dissipative critical systems a different universality class from the one of the equilibrium Mott insulator-superfluid transition.

However, in order to study and classify this critical quantum regime beyond a mean-field analysis, we need to resort to a novel quantum dynamical field theory for driven open systems, which is the core of this work, and which extends Keldysh semi-classical methods employed previously to inspect classical non-equilibrium criticality.
\\

\subsection{Microscopic driven-dissipative action}
\label{DQFT}

In order to develop the quantum dynamical field theory at inspection in this work, we convert the evolution encoded in the quantum master equation,  \eqref{master} into a Keldysh partition function \cite{kamenevbook,Altland/Simons}, in view of renormalization group applications.
The procedure is detailed in a number of works \cite{Sieberer2014, Sieberer2015, Gorshkov2016}, and we will not repeat here all the technical steps, although we outline its main logic. First, it is customary to introduce the Nambu spinor of the complex field, $\hat{\phi}(x)$, on the forward ($+$) and backward ($-$) branch of the (Keldysh) closed time contour \cite{kamenevbook,Altland/Simons}, ${\Phi}_{\sigma}=(\phi_\sigma, {\phi_\sigma}^*)^T$, with $\sigma=\pm$. The action $S$, occurring in the Keldysh partition function, is composed of a Hamiltonian contribution, $S_H$ (roughly speaking a real part), and a dissipative contribution, $S_D$ (imaginary part), which encodes the Lindbladian terms of Eq. \eqref{master}. In general, operators which act on $\hat{\rho}$, from right or left, corresponds to fields which appear in $S$, on the upper ($\sigma=+$) or lower ($\sigma=-$) branch of the Keldysh contour \cite{Sieberer2014, Sieberer2015, Gorshkov2016}. 
For practical convenience, we then express the driven-dissipative microscopic action, $S=S_H+S_D$, associated to the Lindblad dynamics \eqref{master},  in terms of the so-called classical and quantum fields of the Keldysh formalism, 
\be\label{classquant}
\phi_{c,q}=\frac{1}{\sqrt{2}}(\phi^+\pm\phi^-).
\ee
In this basis, the classical field, $\phi_c$, can condense, in the sense that its expectation value in the mean-field BEC condensate phase, yields $\langle\phi_c\rangle=\phi_0$.
On the other hand,  $\phi_q$ describes on equal footing  both noise and quantum fluctuations, and, by definition, it cannot condense,  $\langle\phi_q\rangle=0$.
Employing the change of variable \eqref{classquant}, the Keldysh action, $S$, acquires the so-called causality structure \cite{kamenevbook}: The spectral properties of the system obtains from the retarded (advanced) quadratic part of the Keldysh action $P^{(R,A)}$, which is the inverse of the retarded (advanced) Green's function, $iG^R=\langle\phi_c^*\phi_q\rangle$, while the noise level in the system can be read from the Keldysh component, $P^K$, which relates to the inverse of the Keldysh Green's function, $iG^K=\langle\phi_c^*\phi_c\rangle$. The absence of terms containing only classical fields is a consequence of the conservation of probability \cite{kamenevbook}, and then the action, $S$, in the $c,q$ basis acquires the form


 \begin{widetext}
\be\label{kinetic}
S= \int_{t,x} \left\{\left( \phi_c^{*},\phi_q^{*} \right) \begin{pmatrix}
    0 & P^A \\
   P^R & P^K
  \end{pmatrix}
  \begin{pmatrix}
    \phi_c \\ \phi_q
  \end{pmatrix}+2i\gamma_t\phi^*_c\phi_c\phi^*_q\phi_q 
-  2[(\lambda+i\gamma_t)(\phi_c^{*2}\phi_c\phi_q+\phi_q^{*2}\phi_q\phi_c)+c.c.] \right\}.
\ee
\end{widetext}
The retarded inverse propagator 
\be
\bar{P}^R=(\bar{P}^A)^\dag= i \partial_t + ( {{K}}_R - iK_I ) \partial^2_x + i \chi,
\ee
contains the parameter, $\chi$, which controls the distance from the condensation transition in three dimensions, as we discussed in the previous subsection. Analogously to the fine tuning of the mass of an equilibrium $\phi^4$-theory, the fine tuning of $\chi$ to zero, sets the system to criticality, and for this reason it plays in the following the role of a spectral gap. 

On the other hand, the noise level, 
\be\label{noisequadratic}
P^K=i(\gamma-2\gamma_d\partial^2_x),
\ee
contains both the flat and the diffusion Markov components, induced respectively by the Lindblad operators $L_{a}$, with $a=p,l$ (recall that $\gamma=(\gamma_p+\gamma_l)/2$), and by $L_d$. The diffusion noise introduces a structural relation between the retarded and Keldysh sectors: in both of them a homogeneous constant term is accompanied by a Laplacian. 
This structural relation will play a crucial role in finding a quantum critical scaling regime for the one-dimensional diffusive Bose gas. 

It is also important to stress that in Eq. \eqref{kinetic}  coherent and dissipative processes are now mathematically treated on the same footing; in contrast,  in the Lindblad master equation, \eqref{master}, they are respectively included in $H$ and $\mathcal{L}$. Indeed, after encoding  Lindblad dynamics on the Keldysh contour (see Refs. \onlinecite{Sieberer2015, Gorshkov2016} for further details on the procedure),  coherent couplings are associated to the real part of $S$, while the dissipative ones contribute to its imaginary part.  This is in particular evident looking at quartic vertices, where the real coupling, $\lambda$, associated to coherent density-density scattering comes together with the imaginary coupling, $i\gamma_t$, which is the rate of two-body losses. We can give further structure to the interaction part of the action, noticing that operators odd in the quantum  field, $\phi_q$, are responsible for the physics  associated to a quartic $\phi^4$ potential 
\be
(\lambda+i\gamma_t)\phi_c^{*2}\phi_c\phi_q
\ee
and for the additional fluctuations on the top of it, which are insignificant in the semiclassical limit \cite{Sieberer2014},
\be
(\lambda+i\gamma_t)\phi_q^{*2}\phi_q\phi_c.
\ee
On the other hand, the term even in the quantum fields
\be
i\gamma_t\phi^*_c\phi_c\phi^*_q\phi_q
\ee
represents multiplicative noise induced by two-body losses processes -- a higher order version of the Markov noise induced by one body loss/pump processes. 

Before concluding this subsection, we comment on the poles of the $2\times2$ matrix Green's function, $G(Q)$, as a function of frequency and spatial momentum, $Q=(\omega, q)$:

\be\label{green}
\begin{pmatrix}
    G^K(Q) & G^R(Q) \\
   G^A(Q) & 0
  \end{pmatrix}.
\ee

\begin{figure}[t!]\centering
\includegraphics[width=7.0cm]
{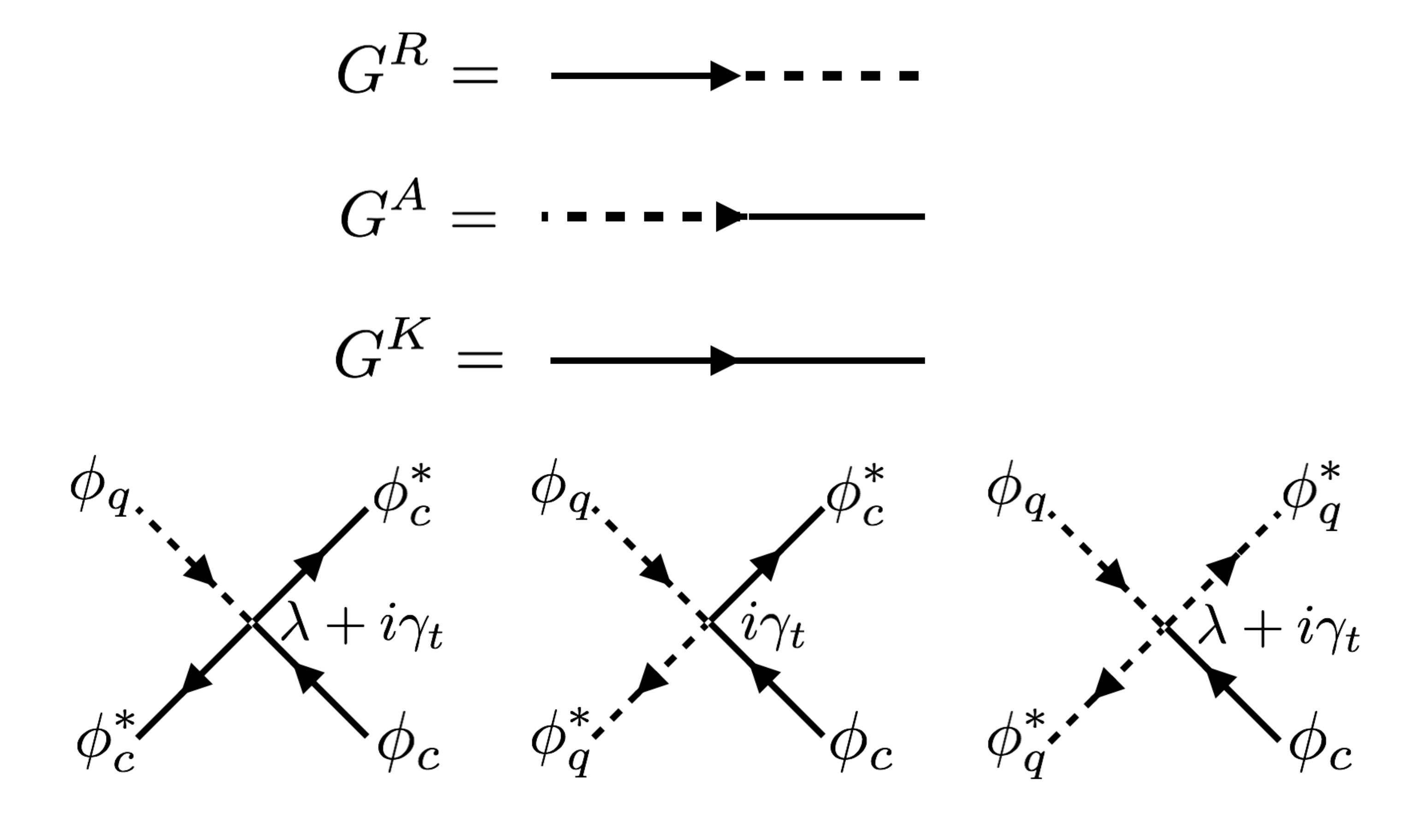}
 \caption{(Color online) Diagrammatic representation of the retarded, advanced and Keldysh Green's functions (propagators), and of the vertices of the action \eqref{kinetic}.
}
\label{Greenfig}
\end{figure}

In Eq. \eqref{green}, $G^R(Q)=1/P^R(Q)$, $G^A(Q)=1/P^A(Q)$ and $G^K(Q)=-P^K(Q)/(P^R(Q)P^A(Q))$ (see Fig. \ref{Greenfig} for a diagrammatic representation ).

The zeros of $G(Q)$ can come only from the zeros of the spectral components $P^{(R/A)}(Q)$, since $P^K(Q)$ enters as a multiplicative factor in $G^K(Q)$. For instance, we can read from the condition $P^R(Q)=0$, the dispersion relation
$\omega=(K_R-iK_I)q^2-i\chi$,
which determines at a mean field level the dynamical critical exponent, $z=2$. In the spontaneously symmetry broken phase ($\rho_0\neq0$), the following complex dispersion relation of long-wavelengths modes is found,
\be\label{Bog}
\omega_{1,2}(q)=-iK_Iq^2-i\chi\rho_0\pm\sqrt{K_Rq^2(K_Rq^2+2\lambda\rho_0)-(\chi\rho_0)^2}
\ee
after rewriting  the complex fields in Eq. \eqref{kinetic} in terms of  real (or Nambu) fields\cite{Sieberer2014}. Eq. \eqref{Bog}
reduces to a diffusive Goldstone mode in the limit $q\to0$, which further confirms that the dynamical critical exponent is $z=2$. This property is a consequence of the spontaneous breaking of $U(1)$ invariance, and persists even when fluctuations are included.  While a condensation regime sets naturally for a three dimensional driven Bose gas, we now clarify how this can happen for an open one-dimensional bosonic system by virtue of the diffusion Markov noise.  

\subsection{Markov diffusion and quantum scaling out-of-equilibrium}
We first qualitatively explain  the role of the diffusion Markov noise in order to achieve a quantum scaling regime in a driven-dissipative Bose gas.

After Fourier transformation, the noise component of the quadratic action, Eq. \eqref{noisequadratic}, contains a momentum dependent term
\be
P^K(q)=i(\gamma+2\gamma_dq^2).
\ee
The importance of the momentum dependent part of the noise, can be understood focusing on  a regime of strong diffusion $\gamma\ll2{\gamma}_d q^2$ (achieved formally in the limit of weak constant Markov noise $\gamma=\gamma_l+\gamma_p\to0$), where a  critical behaviour analogous to zero temperature quantum criticality can settle. Specifically, in a system with dynamical critical exponent, $z=2$ (like the  driven open Bose gas considered here), such diffusion noise can realise, at criticality, a scaling regime analogous to the scaling behaviour of a critical Bose gas coupled to a zero-temperature ohmic  bath of harmonic oscillators, where the noise level has the expression, $P^K_{eq}\sim |\omega|$. As the non-Markovian equilibrium noise, $P^K_{eq}(\omega)$, is responsible for an equilibrium \emph{quantum} phase transition at zero temperature, analogously the diffusion noise, $P^K_{neq}(q)\sim\gamma_dq^2$, is responsible for a non-equilibrium \emph{quantum} phase transition in the driven Bose gas, in the sense that they scale identically at the level of canonical power counting $P^K_{eq/neq}\sim q^2$ (recall again that $\omega\sim q^z$, $z=2$, see Eq. \eqref{Bog} and related discussion). This discussion technically defines the analogy between equilibrium and non-equilibrium quantum criticality; below we show indeed that coherent quantum dynamics persists to large scales in our setup, unlike the equilibrium or non-equilibrium classical criticality (cf. Ref. \onlinecite{Sieberer2014}). According to the analogy with zero temperature quantum criticality, fluctuations relevant in  RG sense in the former case will be relevant also in  our setup, which can exhibit, in addition, novel effects dictated by its genuine non-equilibrium driven nature. By virtue of this analogy, we expect also in the driven open Bose gas a dimensional enlargement, $D=d+z$, typical of quantum phase transitions in $d$ dimensions: this circumstance  allows for condensation in the low dimensional setting considered in this work ($d=1$).

\begin{figure}[t!]\centering
\includegraphics[width=6.5cm]
{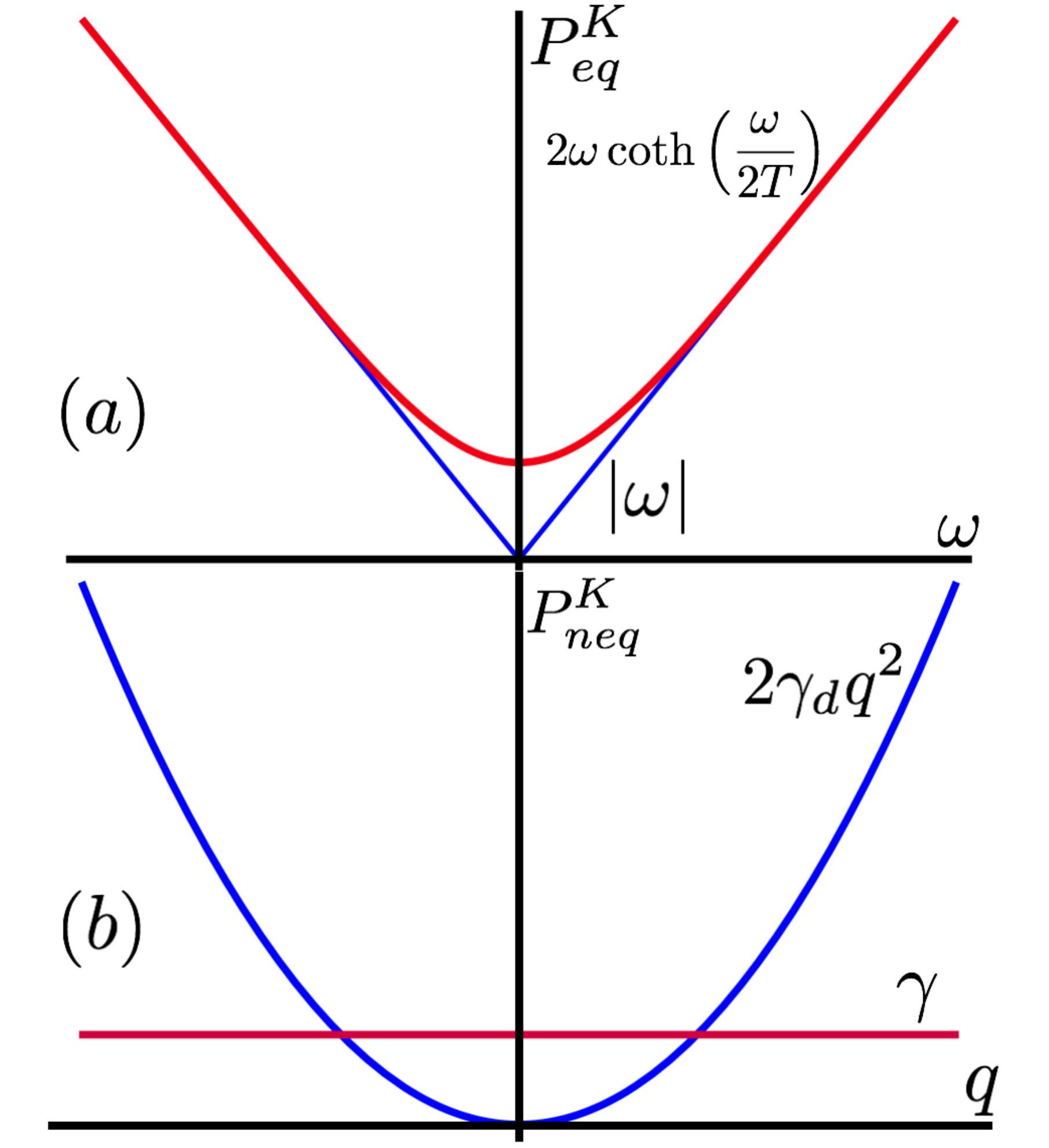}
 \caption{(Color online) 
 A  comparison between the Markovian non-equilibrium noise, $P^K_{neq}$ (panel (b)) and the equilibrium non-Markovian quantum noise, $P^K_{eq}=\lim_{T\rightarrow0}\coth(\frac{\omega}{2T})=|\omega|$, which constitutes the  zero-temperature limit (blue line in panel (a)) of the noise induced by a Caldeira-Leggett  bath (red line in panel (a)).   In particular, in (b) the red straight line indicates that there is a region of low  momenta ($q<\Lambda_M$) where the flat Markovian noise (red line)  dominates over the  diffusion Markov noise (blue line).}
\label{noise}
\end{figure}

Despite their similarity with respect  to a canonical power counting analysis, the quantum diffusion,  $P^K_{neq}\sim \gamma_dq^2$,  is weak at long wavelengths and Markovian at all length scales, while the equilibrium quantum noise, $P_{eq}^K(\omega)\sim|\omega|$, is non-Markovian at all length scales and decays at long times like $P_{eq}^K(t-t')\sim\frac{1}{|t-t'|^2}$. Nevertheless, they are both 'gapless' (i.e. respectively, the mode $\omega=0$ and $q=0$ are decoupled from the noise) around the origin -- a key feature to realise quantum critical scaling. Notice that in  contrast  a flat Markovian noise level ($P^K=i\gamma$, $\gamma_d=0$)  is by definition 'gapped'  around the origin, as the noise level of a high temperature bath ($P^K=i T$).  Fig. \ref{noise} summarises the comparison between $P_{eq}^K$ and  $P_{neq}^K$.

\subsubsection{Quantum canonical scaling}
\label{QCanonical}

We now discuss  the analogy between $P^K_{eq}$ and $P^K_{neq}$ on the basis of canonical power counting arguments.

We start considering  canonical power counting at finite temperature, $T = 1/\beta$. 
Instead of the driven-dissipative  $O(2)$ model, discussed in Sec. \ref{DQFT}, we consider an over-damped $\phi^4$ scalar model which has thermalised with  an ohmic  bath of quantum harmonic oscillators at temperature $T$, known as Caldeira-Leggett model \cite{caldeira, Breuerbook} (see  Sec. 8.6 of the book in Ref. \onlinecite{kamenevbook}, for a detailed field theoretical analysis of the related phase transition).  

In this case, the quadratic part of the action has to be supplemented by
the following terms induced by the bath (where $\phi_{\nu}$ with $\nu=c,q$, denotes the Fourier transform of the classical and quantum fields here),
\begin{equation}
  \label{eq:10}
  S_{\mathrm{reg}} = \int_Q \left( {\phi}_c^{*}, {\phi}_q^{*} \right)
  \begin{pmatrix}
    0 & + i \delta \omega \\
    -i \delta \omega & i  \delta \omega \coth(\beta \omega/2)
  \end{pmatrix}
  \begin{pmatrix}
   {\phi}_c \\ {\phi}_q
  \end{pmatrix},
\end{equation}
which automatically fulfils the fluctuation-dissipation theorem. The parameter $\delta$ is a strong dissipation coupling induced by the bath of oscillators, and it determines the overdamped nature of the system,  setting also in this case $z=2$.

We
distinguish for the distribution function $\coth(\beta \omega/2)$ the case of
finite temperature where the behavior at low frequencies $\omega \ll T$ is
$\coth(\beta \omega/2) \sim 2T/\omega$, whereas for $T \to 0$ we have
$\coth(\beta \omega/2) \to \mathrm{sgn}(\omega)$. This implies that in the latter case, the canonical scaling dimension of the noise level is $[P^K]=2$, while in the former we have $[P^K]=0$. The notation $[...]$ indicates for a given term in the action its canonical dimension  measured in units of momentum (for instance, $[g]=n$ means $g\sim q^n$). 

Recalling that the action must be dimensionless, and applying this criterion to the Gaussian action Eq. \eqref{eq:10}, we find for the canonical dimensions of classical and quantum fields, at $T=0$
\be
  d_c =  d_q =  \frac{d}{2},
\ee
while at high temperatures ($\omega\ll 2T$), 
\begin{equation}
  \label{eq:11}
  d_c = \frac{d - 2}{2}, \quad d_q  = \frac{d + 2}{2}.
\end{equation}

The fact that the canonical scaling dimension of the Keldysh component is $[P^K]=2$ at $T=0$, shows that the power counting is the same for the driven model in Eq. \eqref{kinetic} in the presence of diffusion noise, $P^K_{neq}$,  for  $\gamma= 0$ but $\gamma_d \neq 0$  (in the entire analysis we set
$[\gamma] = [\gamma_d] = [\delta]=0$). On the other hand, when $\gamma
\neq 0$, we recover the canonical scaling $[P^K]=0$, of a high temperature noise level.

Analogously to equilibrium quantum phase transitions, the novel non-equilibrium quantum critical regime of driven open Bose gases, can be accessed under the double fine tuning of two parameters: it is required a strong diffusion Markov noise, which is formally achieved through the fine tuning of the Keldysh mass to zero, $\gamma\rightarrow0$ (or $\gamma_l$,$\gamma_p\rightarrow0$), and  a fine tuning of the spectral mass to zero, $\chi\rightarrow0$, ($\gamma_l\rightarrow\gamma_p$). In the same way, an equilibrium quantum critical point is reached after  fine tuning its spectral gap, and setting formally the temperature to zero, $T\to0$, since it  always constitutes a relevant perturbation which  destroys quantum criticality.

The double fine tuning $\gamma_l\to\gamma_p\to0$ and $\gamma=\gamma_l+\gamma_p\to0$ is never perfectly achievable, since the flat noise component of the Markov noise ($\gamma$) can never be completely erased in realistic experimental situations, as fine tuning the temperature $T\to0$ is not an asymptotically achievable condition in equilibrium quantum phase transitions.  However, a situation with a non-equilibrium quantum regime, with $\gamma \ll \gamma_d
q^2$ in some momentum range, is conceivable, and this triggers the question whether
there is an interacting (i.e., non-Gaussian) fixed point for the RG flow of the action \eqref{kinetic}. This fixed point  governs
 the RG flow in the quantum regime, leading to non-trivial
scaling behavior of correlation functions, before the flow ultimately enters the thermal scaling regime below the
momentum scale, $\Lambda_{\mathrm{M}} \simeq \sqrt{\gamma/\gamma_d}$, which in the following we refer to as Markov scale. This behaviour is  analogous to finite
temperature quantum phase transitions, where  quantum scaling is destroyed for distances larger than the de-Broglie wavelength, $L_{dB}\sim 1/T^{1/z}$. However, it is important to stress that the  mean-field estimate $\Lambda_M\simeq\sqrt{{\gamma}/{\gamma_d}}$,  is premature;  the Markovian noise level, $\gamma(k)$, is in general dependent on the RG running scale $k$, and it acquires corrections during the RG flow;  its value can be  renormalized by loop corrections. This  requires an estimate of $\Lambda_M$ from the effective infrared action, which will be presented in Section \ref{Sec:extent}, together with the estimate of the upper momentum  bound for non-trivial scaling behaviour, which we refer to as Ginzburg scale, $\Lambda_G$ , see Ref. \onlinecite{Amit/Martin-Mayor}.





In the parameter regime of strong diffusion Markov noise of the  driven open gas, spectral and Keldysh components of the Gaussian action, Eq. \eqref{kinetic}, scale with the same power of momentum, $P^{R/A/K}(q)\sim q^2$, and  accordingly the classical and quantum fields have the same canonical dimension, $[\phi_c]=[\phi_q]=d/2$ (in  $d$ spatial dimensions). As we have discussed above, this occurs also for quantum phase transitions of open systems at $T=0$.
However, since the canonical scaling dimension of quartic couplings is $[g]=2-d$,  any quartic combination of fields which respects  $U(1)$ symmetry is generated in the course of renormalization (irrespective of the number of quantum fields involved in each quartic vertex), and needs to be  incorporated in a renormalization group description for the driven-diffusive Bose gas. For instance, the operators  $\phi^{*2}_c\phi^{2}_q$ or $\phi^{*2}_q\phi^{2}_q$ must be included because they are RG relevant, even if not present in the microscopic description of Eq. \eqref{kinetic}. 

As a last important remark, we notice that, since  the noise level is scaling quadratically with momentum, we have  $[\gamma]=2$ in the quantum scaling regime, showing that $\gamma$, as a temperature, is a relevant perturbation at a quantum critical point. This  supports the intuition that  the quantum scaling regime can be seen as  effectively tuning to zero the flat Markovian noise level in the momentum window, $\Lambda_M<q<\Lambda_G$.

The upper critical dimension of the system, $D_c=2$ (recall that $[g]=2-d$), in the presence of diffusion Markov noise, is the reason why we are considering the  driven open Bose gas in  one dimension. 
As previously anticipated, since the regime of strong Markov diffusion is realising an analog of quantum critical behaviour, a dimensional enhancement of the phase space is expected, and a regime of condensation can occur even in one spatial dimension, since the effective space dimensionality is $D=d+z=3$, in the momentum range where the diffusion Markov noise dominates over the flat Markov noise level ($\gamma_dq^2\gg\gamma$). On the other hand, when $\gamma\gg\gamma_d q^2$ such  effective dimensional enhancement does not hold anymore since the leading noise effect has a thermal like character ($P^K\sim\gamma$ is analogous to a high temperature Caldeira-Leggett noise), and, like in the dimensional crossover occurring in equilibrium finite temperature quantum phase transitions, the effective dimensionality of the system shrinks to its spatial dimensionality ($D\to d$).  In our setup this would correspond to a dimensional shrinking to $D=1$ induced by the flat Markovian noise, where  no continuous symmetry breaking or quantum criticality can occur. In terms of length scales, this mechanism would occur for length scales $l>\Lambda^{-1}_M$: below this scale, no criticality can be present and correlations become exponentially decaying (again in full analogy to the shape of correlation functions for $l>\Lambda_{dB}$, in a finite $T$ quantum phase transition).

Before moving to the next paragraph, we notice that in a more standard closed one dimensional system, a linearly dispersing sound mode, characteristic of a  Luttinger liquid description, is expected, and accordingly, $z=1$, with an effective phase space of $D=z+1 = 2$, which prevents condensation to occur. However, the linear dispersion relation can be attributed to  particle number conservation, an ingredient which is missing in our driven open setup, and which rules out sound waves in favour of a diffusive dispersion relation (see Eq. \eqref{Bog}).



\subsubsection{Semi-classical scaling}
Before concluding this section, we briefly recall the properties of the scaling regime associated to the  Bose gas in the presence of flat Markov noise ($P^K=i\gamma$, with $\gamma_d=0$).

In  contrast to what has been discussed for the one dimensional driven open gas, the canonical scaling in the presence of a homogeneous Markovian noise ($\gamma_d=0$), implies that  operators non-linear in the quantum fields are irrelevant, that the upper critical dimension is $D_c=4$, and accordingly that the only relevant vertex is the so-called classical coupling, which supports vertices with three classical fields and one quantum field, ${\phi_c^{*2}}\phi_c\phi_q$. The canonical critical scaling regime is in this case the same of a classical finite temperature phase transition (as discussed above), i.e. the constant Markovian noise  does not scale at the canonical level, $\gamma\sim q^0$, and the spectral part and noise level of the quadratic Keldysh action scale differently, $P^{R/A}\sim q^2$, $P^K\sim q^0$. In a nutshell, in order to capture the three-dimensional driven-dissipative Bose transition, it is sufficient to resort to a semi-classical approximation: the Hamiltonian part of the action contains only quartic operators linear in $\phi_q$, while the noise term is just quadratic in $\phi_q$.

The open Bose gas coupled to diffusion Markov noise realises a concrete situation where it is necessary to move beyond such semi-classical approximation,  studied in full detail in Refs. \cite{Sieberer2013, Sieberer2014} and whose non-equilibrium universality class constitutes a benchmark for the forthcoming results. %

To summarise, in this work we are looking for new scaling solutions of the Keldysh action, where spectral gap and noise level, scale both quadratically $\chi\sim q^2$, $\gamma\sim q^2$, and search for the  interacting, i.e. non-Gaussian, fixed point of the associated renormalization group flow. Indeed, it is the functional renormalization of the  quantum dynamical field theory of next section which achieves this goal.

\subsubsection{Comparison with Model B of Hohenberg-Halperin classification}
 
We conclude this Section with a comment for the reader familiar with Hohenberg-Halperin classification of equilibrium critical dynamics. 

A diffusion noise is also present in the Model B of Hohenberg-Halperin classification \cite{tauberbook, HHRev} -- a purely relaxational field theory whose dynamics conserves the order parameter, $\phi(x,t)$ ($x$ is the $d$-dimensional space coordinate in this paragraph). The equation of motion for $\phi(x,t)$ is provided by the stochastic equation of motion
\be
\frac{\partial \phi(x,t)}{\partial t}=D\nabla^2\frac{\delta \mathcal{H}(\phi)}{\delta \phi(x,t)}+\xi(x,t),
\ee 
equipped with a random noise of zero mean ($\langle\xi(x,t)\rangle=0$), whence 
\be
\langle \xi(x,t) \xi(x, t')\rangle=-2DT\nabla_x^2\delta(x-x')\delta(t-t').
\ee
The Hamiltonian, $ \mathcal{H}(\phi)$, is the one of the $\phi^4$ model, and the generator of dynamics carries an overall Laplacian term, in order to entail the conservation of $\phi$. This overall derivative determines a dynamical critical exponent, $z=4$, and, accordingly, a canonical power counting as in a classical problem. In contrast, in our problem the overall Laplace term is missing (the dynamics does not conserve the order parameter), $z=2$, and the power counting of a quantum problem can therefore be achieved, as discussed above. 

This implies, as already stressed above, that quantum effects beyond semi-classical truncations and multiplicative noise effects,  are relevant in a RG sense at the critical point of the one-dimensional driven  Bose gas, and so the universality class is expected to be different from the one of model B.
Besides, the fine tuning to reach criticality in model B involves only the spectral gap, while in our problem both spectral gap and noise requires to be tuned.

\subsection{Diffusion Markov noise from microcavity arrays}
\label{appA}

We now start a digression on an experimental set-up where the diffusion Markov noise can be realised. 

The focus is on the engineering of the Lindblad operator, $L_d(x)$, which can be realised through a quantum optics architecture in one spatial dimension, particularising ideas already present in Ref. \onlinecite{Marcos12}. \\

We start with a simple open quantum system \cite{Breuerbook}:
a superconducting qubit ($\sigma^z$) coupled to two microwave bosonic excitations, living in two close optical cavities, whose master equations reads

\begin{equation}
\begin{split}
\dot{\rho}&=-i[H,\rho]+\Gamma\Big(L\rho L^\dag-\frac{1}{2}\{L^\dag L,\rho \}\Big),\\
H&=H_0+H_{R},\\
H_0&=\omega_cc_1^\dag c_1+\omega_cc_2^\dag c_2+\omega_0\sigma^z,\\
H_{R}&=\Omega(\sigma^+c_a+\sigma^-c_a^\dag),\quad L=\sigma^-.
\end{split}
\end{equation}

The Lindblad operator $L$ accounts for spontaneous decay of the qubit with rate $\Gamma$, while bosonic excitations are created in the two cavities, by the raising  bosonic operators $c_1^\dag$ and $c_2^\dag$. The qubit coherently couples to the anti-symmetric superposition of bosonic degrees of freedom, $c_a=c_1-c_2$, through a Rabi term, $H_R$, of frequency $\Omega$.
In the limit $\Gamma\gg \Omega$,  the dynamics of the qubit degrees of freedom is faster than the bosonic one, and it can be adiabatically eliminated through a second-order perturbative projection operator method (see e.g. Section 10.2 of \onlinecite{Breuerbook}):  On time scales $t\gg1/\Gamma$, the system relaxes into the state,  $\rho(t)\simeq\rho_c(t)\otimes\rho_q^0$, where $\rho_q$ is the zero-temperature reduced density matrix of the qubit and $\rho_c(t)$ is the reduced density matrix of the bosonic degrees of freedom. The qubit  is assumed in a zero temperature state (i.e. $\omega_c\gg T$, where $T$ is the environmental temperature), which is a realistic assumption for current quantum circuit realisations.  

The resulting effective master equation for the reduced cavity density operator, $\rho_c$, can then be  written as

\begin{equation}
\partial_t\rho_c=-i[H_C+H_{LS},\rho_c]+\gamma_d\big(c_a\rho_cc_a^\dag-1/2\{{c_a^\dag c_a,\rho_c}\}\big),
\end{equation}
where 
\begin{equation}
\gamma'_d=\frac{\Omega^2\Gamma}{[(\omega_0-\omega_c)^2+(\Gamma/2)^2]},
\end{equation} 
and
\begin{equation}
H_{LS}=\frac{g^2}{2}\frac{(\omega_0-\omega_c)}{\left(\frac{\Gamma}{2}\right)^2+(\omega_0-\omega_c)^2}c_a^\dag c_a,
\end{equation}
is the Lamb Shift Hamiltonian. \\

\begin{widetext}

\begin{figure}[t!]\centering
\includegraphics[width=15.5cm]
{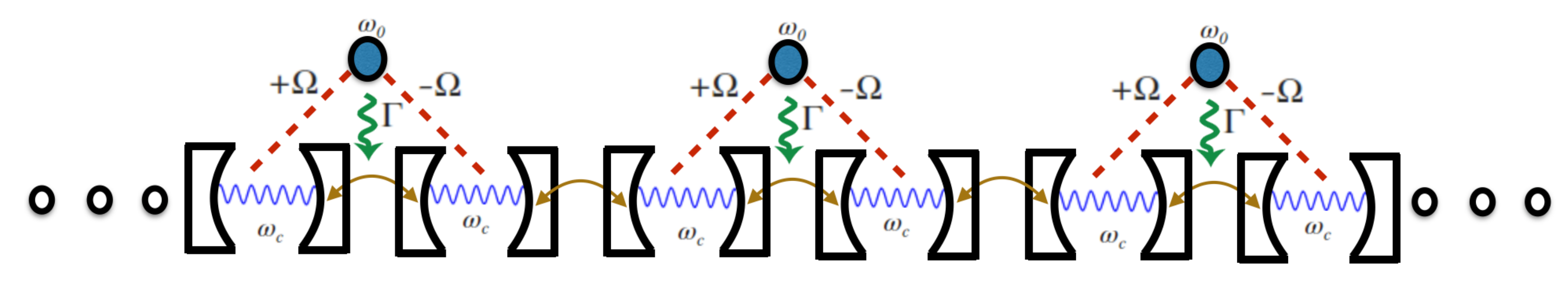}
 \caption{(Color online) 
 The quantum optics architecture employed to engineer  the Lindblad operator, $L_d=\partial_x\phi(x)$. Yellow curved lines stand for hopping processes among  bosonic excitations inside the cavities.}
\label{micro}
\end{figure}

\end{widetext}

Generalizing this setup to a many-body system of $i=1,2, ..., N$ qubits and $2N$ cavities (as portrayed in Fig. \ref{micro}), and taking the continuum limit of the system, we can convert the antisymmetric combination of the cavity modes, $c_a\sim c_i-c_{i+1}$, into a spatial gradient, $c_a\to\alpha^2\partial_x c(x)$, which corresponds in a field theoretical language to the Lindblad operator, $L_d=\partial_x\phi(x)$ (discussed already in Sec. \ref{microscop}), with coupling $\gamma_d=\alpha^3\gamma'_d$ ($\alpha$ is the lattice spacing of the many-body cavity model). 

\section{Quantum dynamical field theory and Functional Renormalization Group} 
\label{FRGSec}

Our goal is now to dress the microscopic coefficients of the quantum Lindblad equation Eq. \eqref{master} with renormalization group corrections. We employ a functional renormalization group (FRG) approach, which has been already successfully applied to the driven-dissipative condensation transition of a three dimensional Bose gas \cite{Sieberer2014, Sieberer2015}, as well as to other non-equilibrium situations~\cite{Canet2004,Canet2005,Canet2010,Gezzi2007,Jakobs2007,Karrasch2010, gasenzer08:_towar,berges09:_nonth, Kessler2010}.

We will not repeat here how to adapt  the FRG methods to the study of a Keldysh action for a driven open system. We instead briefly summarise the main steps of the procedure; the interested reader can find details  in Refs. \onlinecite{Sieberer2014, Sieberer2015}.

\subsection{Flow of the effective action}
\label{FRGWettSec}

Our starting point is the Wetterich equation\cite{berges02:_nonper} , a functional differential equation for the effective action $\Gamma_{k}$ -- associated to the Keldysh microscopic action \eqref{kinetic},  
\begin{equation}\label{Wetterich}
\partial_k\Gamma_k=\frac{i}{2} \operatorname{Tr} \Big[(\Gamma_k^{(2)}+R_k)^{-1}\partial_kR_k\Big],
\end{equation}
where the trace operation, $ \operatorname{Tr}$, denotes summation over external indices as well as summation over frequencies and momenta. $\Gamma_k^{(2)}$ is the second variational derivative with respect to classical and quantum fields of the effective action, $\Gamma_k$, which compared to the microscopic Keldysh action, Eq. \eqref{kinetic}, has the advantage to incorporate at the same time both statistical and quantum fluctuations \cite{berges02:_nonper}, but which constitutes an equivalent description to the Keldysh path integral formulation discussed in Sec. \ref{DQFT}. However, since $\Gamma_k$ obeys the flow equation \eqref{Wetterich}, it constitutes a more viable tool to perform our renormalization group program. 
In Eq. \eqref{Wetterich} we adopted Litim's cutoff   \cite{litim00:_optim} 

\be\label{litim}
R_k(q)=K(q^2-k^2)\theta(k^2-q^2), 
\ee
%
which does not affect the frequency dependence of the propagator, and it acts adding, for momenta $q<k$, an effective mass proportional to the square of the running scale $\sim k^2$; for instance, the retarded propagator in momentum space reads now
\be
P^R(Q)=\omega-K^*q^2+R^*_k(q),
\ee
where $K\equiv K_R+iK_I$ is the complex kinetic coefficient.
The inclusion of the regulator, $R_k(q)$,  protects the diagrammatic corrections of the theory from infrared divergences coming from critical modes; on the other hand, the presence of the regulator in Eq. \eqref{Wetterich} prevents the development of ultraviolet divergences. This combined effect makes the equation for the functional flow equation, Eq. \eqref{Wetterich}, a powerful tool to renormalize the effective action of the theory:
Eq. \eqref{Wetterich} interpolates from the microscopic Markovian action \eqref{kinetic}, at the ultraviolet scale, $k\simeq k_{UV}$, to the infrared, $k\to0$, effective action. In the course of the functional RG flow, couplings are dressed by loop corrections which determines their associated beta functions: specifically,
in order to convert the functional differential equation, Eq. \eqref{Wetterich}, into a system of non-linear differential equations (beta functions) for the flow of the couplings, we provide an ansatz for $\Gamma_k$ based on the quantum canonical power counting introduced in  Section \ref{QCanonical}. 

\subsection{Quantum dynamical field theory}
\label{introSQ}

We make an ansatz for $\Gamma_k$ capable to capture the Wilson-Fisher (i.e. interacting) fixed point associated to the quantum scaling regime of the Keldysh action, discussed in Section \ref{QCanonical}: we refer to this ansatz, as  \emph{quantum action},  $S_{Q,k}\equiv S_{0,k}+S_{I,k}$, since the interaction action, $S_{I,k}$, contains the whole set of operators relevant in the quantum scaling regime. $S_{Q,k}$ defines the quantum dynamical field theory developed in this work.

\begin{table}[t!]
\centering
\begin{tabular}{|l|l|}
\hline
Couplings & Microscopic values  \\ \hline
$Z$ wave-function renorm. coefficient  & $Z|_{k_{UV}}=1$  \\ \hline
 $\bar{K}_R$ coherent propagation  & $\bar{K}_R|_{k_{UV}}=\frac{1}{2m}$  \\ \hline
 $\bar{K}_I$ diffusion coefficient  & $\bar{K}_I|_{k_{UV}}=\gamma_d$  \\ \hline
 ${\bar{\gamma}}_d$ Markov diffusion noise coefficient & ${\bar{\gamma}}_d|_{k_{UV}}=\gamma_d$  \\ \hline
 $\bar{\chi}$ spectral gap (retarded mass) & $\bar{\chi}|_{k_{UV}}=\frac{\gamma_l-\gamma_p}{2}$  \\ \hline
 $\bar{\gamma}$ Markovian flat noise (Keldysh mass)  & $\bar{\gamma}|_{k_{UV}}=\gamma_l+\gamma_p$  \\ \hline
 $\bar{\lambda}_c$ classical coherent vertex &  $\bar{\lambda}_c|_{k_{UV}}=\lambda$   \\ \hline
$\bar{\kappa}_c$ classical dissipative vertex &  $\bar{\kappa}_c|_{k_{UV}}=\gamma_t$\\ \hline
 $\bar{\lambda}_q$ quantum coherent vertex &  $\bar{\lambda}_q|_{k_{UV}}=\lambda$   \\ \hline
$\bar{\kappa}_q$ quantum dissipative vertex &  $\bar{\kappa}_q|_{k_{UV}}=\gamma_t$\\ \hline
$\bar{g}_1$ non-multiplicative Markovian noise &  $\bar{g}_1|_{k_{UV}}=2\gamma_t$\\ \hline
$\bar{g}_2$ non-gaussian Markovian noise &  $\bar{g}_2|_{k_{UV}}=0$\\ \hline
$\bar{g}_3=\bar{\lambda}_3+i\bar{\kappa}_3$ dephasing jump operator &  $\bar{g}_3|_{k_{UV}}=0$\\ \hline
\end{tabular}\caption{Full list of the couplings occurring in $S_Q$. In the microscopic driven open Bose gas with action \eqref{kinetic}, we used instead the following labels: $m$ for the mass of bosons, $\lambda$ for the strength of two-body collisions, $\gamma_d$ for the rate of diffusion, $\gamma_{l,p}$ for the one body loss/pump rates, $\gamma_t$ for the two-body loss rate.}
\label{tabmicro}
\end{table}
$S_{0,k}$ is  identical to the quadratic part of Eq.~\eqref{kinetic}, provided a complex wave-function renormalization coefficient, $Z$ ($Z\equiv Z_R+iZ_I$), has been introduced in the inverse retarded/advanced propagator, 
\be\label{ret}
\bar{P}^R=iZ^*\partial_t+\bar{K}^*\partial^2_x,
\ee 
while for the noise level of the quadratic action we have
\be
\bar{P}^K=i(\bar{\gamma}-2\bar{\gamma}_d\partial^2_x).
\ee
From now on, we are adding a bar on the top of couplings and fields, and we will refer in the following to them as \emph{bare} couplings and fields. Later, they will be rescaled through the factor $Z$, and we will define accordingly a set of renormalized (by $Z$) couplings and fields. 
It should be noted that in Eq. \eqref{ret} the spectral mass, $\bar{\chi}$ has not been introduced: Since we are going to expand the action, $S_{Q,k}$, around a stationary condensate solution ($\bar{\phi}_c=\bar{\phi}^*_c=\sqrt{\phi_0}$ and $\bar{\phi}_q=\bar{\phi}^*_q=0$),  the retarded mass will be effectively generated by the condensate. This can be  realised immediately, displaying the ansatz for the homogeneous part of the action which contains quartic terms, $S_I$, \begin{equation}\label{int}
\begin{split}
S_I&=S_h+S_a,\\
S_{h}&=- \int_{\mathbf{x},t}  \frac{1}{2}\left[\frac{\partial{\mathcal{\bar{U}}}_c}{\partial\bar{\phi}_c}\bar{\phi}_q+ \frac{\partial{\mathcal{\bar{U}}}_c^*}{\partial\bar{\phi}_c^*}\bar{\phi}_q^*+ \frac{\partial{\mathcal{\bar{U}}}_q}{\partial\bar{\phi}_q}\bar{\phi}_c + \frac{\partial{\mathcal{\bar{U}}}_q^*}{\partial\bar{\phi}_q^*}\bar{\phi}_c^*\right],\\
S_{a}&= \int_{\mathbf{x},t}  i\bar{g}_1 (\bar{\phi}_c^{*} \bar{\phi}_c-{\bar{\rho}_0}/{2})\bar{\phi}_q^{*} \bar{\phi}_q + i \bar{g}_2 ( \bar{\phi}_q^{*} \bar{\phi}_q)^2 \\
&- \frac{1}{4} [ \bar{g}_3( \bar{\phi}_c^{*} \bar{\phi}_q )^2 -{\bar{g}_3}^*( \bar{\phi}_c \bar{\phi}_q^* )^2 ].
\end{split}
\end{equation}
$S_h$ and $S_a$ are  the hermitian (odd in the quantum fields, $\bar{\phi}_q$) and anti-hermitian (even in the quantum fields, $\bar{\phi}_q$) parts of the interaction action. The mass term in the retarded sector (to which we refer as retarded mass in the following) is effectively generated by the condensate, $\chi=-\bar{\rho}_0\bar{\kappa}_c/2$, while the flat Keldysh noise level, $\bar{\gamma}$, will be labelled as Keldysh mass, since in the quantum scaling regime it scales quadratically like a spectral gap.

The potentials, ${\mathcal{\bar{U}}}_c=\frac{1}{2}\bar{u}_c(\bar{\phi}_c^*\bar{\phi}_c-\bar{\rho}_0)^2$ and ${\mathcal{\bar{U}}}_q=\frac{1}{2}\bar{u}_q(\bar{\phi}_q^*\bar{\phi}_q)^2$, are respectively supported by the  complex couplings    $\bar{u}_{c,q}\equiv\bar{\lambda}_{c,q} + i \bar{\kappa}_{c,q}$, whose real and imaginary parts  coincide respectively with the interaction strength, $\lambda$, and the two body loss rate   entering the quantum master equation Eq. \eqref{master}, $\bar{\lambda}_c|_{k_{UV}}=\bar{\lambda}_q|_{k_{UV}}=\lambda$, $\bar{\kappa}_c|_{k_{UV}}=\bar{\kappa}_q|_{k_{UV}}=\gamma_t$. 

${\mathcal{\bar{U}}}_c$ is the usual quartic potential responsible for the formation of a condensate phase, while ${\mathcal{\bar{U}}}_q$ is its quantum counterpart, accounting for  quantum fluctuations on  top of the quartic classical potential. Such quantum vertices are irrelevant (in RG sense) in a semi-classical truncation, which includes quartic operators up to linear order in $\bar{\phi}_q$, and which is expected to capture the condensation transition of a three dimensional open Bose gas driven by flat Markovian noise. 

The coupling $\bar{u}_q$ is  the first genuine  ingredient which distinguishes our quantum dynamical field theory from the standard MSRJD equilibrium action or from classical dynamical field theories of the
Hohenberg-Halperin classification \cite{tauberbook}. 

As an important remark, we notice that in the quantum action, $S_{Q,k}$, couplings have been relabelled with different names than the ones adopted from the microscopic action. This has a two-fold purpose: first of all,  in Eq. \eqref{kinetic},  couplings have their microscopic values which can be sensitively altered by RG corrections, and, at the same time, we recall that the couplings in $S_{Q,k}$  are running functions of the momentum scale, $k$. 

\begin{figure}[t!]
\includegraphics[width=3.0cm]
{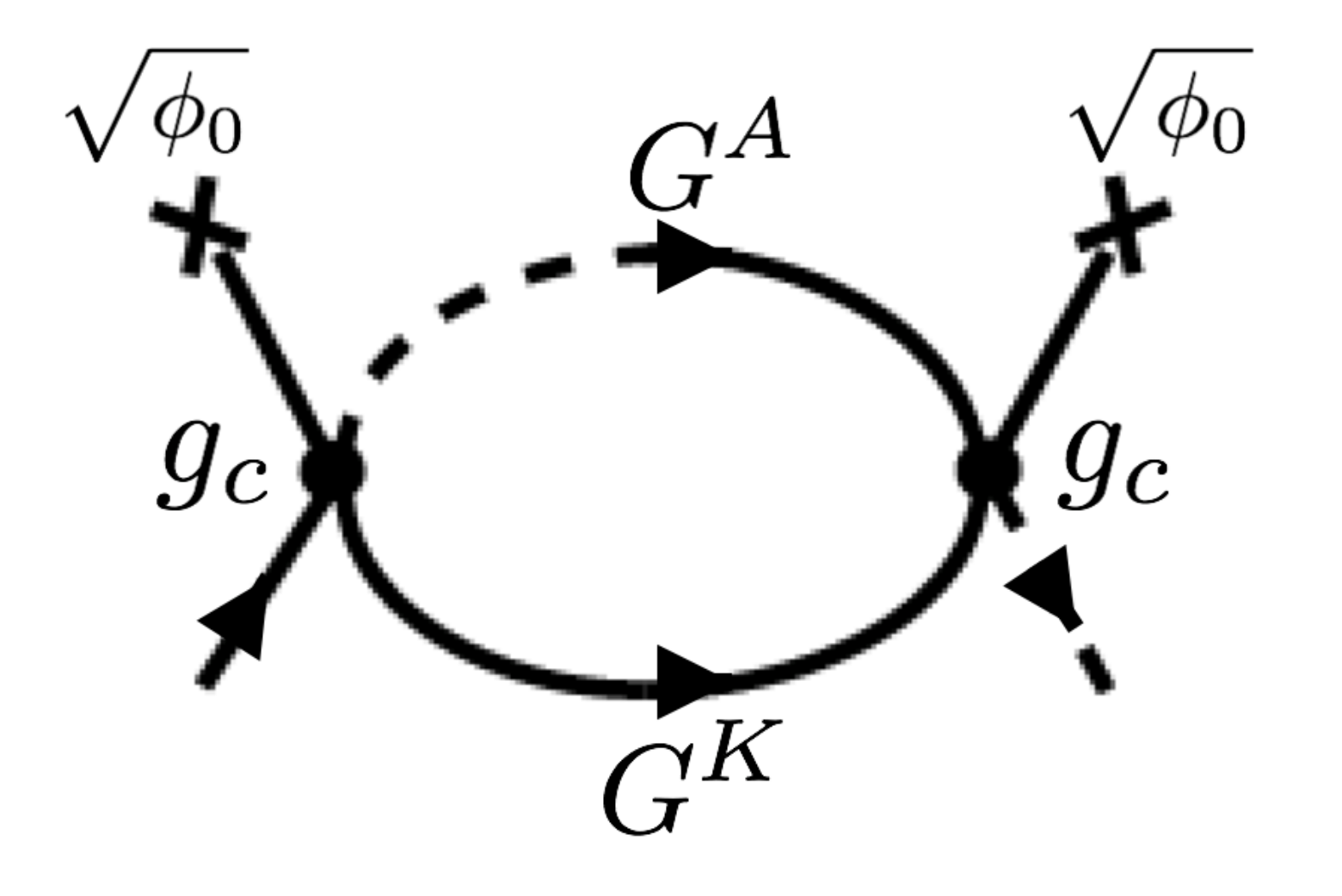}
 \caption{ 
One-loop diagram  renormalizing the momentum and frequency dependent part of the retarded propagator, $G^R$. It can be seen as a two-loop diagram, where an internal Keldysh line has been cut, and the two resulting  free lines have been replaced by the non-vanishing expectation value of the classical field in the ordered phase, 
${\bar{\phi}}_c={\bar{\phi}^*}_c=\sqrt{\phi_0}$. Its frequency and momentum dependence for $q\to0$ and $\omega\to0$ generates the renormalization of  $Z$ and $\bar{K}$, and allows us to  compute their anomalous dimensions. In this diagram, and in others to follow, we drop the over-line  on the couplings.}
\label{cond}
\end{figure}

In Eq. \eqref{int} we introduced the condensate density, $\bar{\rho}_0$, since in  practical calculations we  approach the transition from the ordered phase, taking the limit of the stationary state condensate, $\bar{\rho}_0=\bar{\phi}_c^*\bar{\phi}_c|_{ss}=\bar{\phi}_0^*\bar{\phi}_0\rightarrow0$. In this way, we capture two-loop effects \cite{berges02:_nonper} necessary to extract the full set of critical exponents. Moreover, the presence of a background field introduces a series of one-loop diagrams where external insertions of the classical fields are evaluated on their condensate expectation values, $\bar{\phi}_c=\bar{\phi}^*_c=\sqrt{\phi_0}$; such diagrams   allow us to extract the anomalous dimensions of kinetic coefficients, while keeping  the computations formally at one-loop level, see Fig. \ref{cond}.

We now discuss the noise terms entering $S_a$. The coupling, $\bar{g}_1$, (at the microscale, ${\bar{g}_1}|_{k_{UV}}=2\gamma_t$) constitutes a dressing of the Markovian noise level, $\bar{\gamma}$, and it constitutes a multiplicative noise, which originates when two particles are incoherently drained away from the system, in full analogy to the way in which one-body pump and losses  generate $\bar{\gamma}$.

\begin{figure}
\includegraphics[width=8.2cm]
{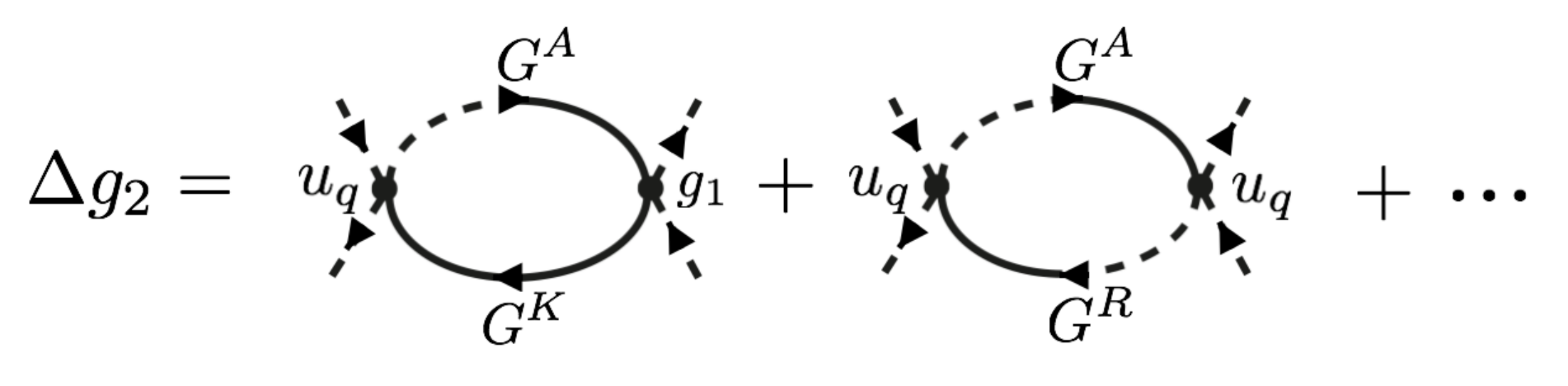}
 \caption{(Color online) 
The picture shows two one-loop diagrams supported by couplings present  at the microscale ($u_q|_{k\sim k_{UV}}=\lambda+i\gamma_t$, $g_1|_{k\sim k_{UV}}=2\gamma_t$), which contribute to effectively generate the non-Markovian noise coupling $g_2$ during RG flow.}
\label{nongaussian}
\end{figure}

Other RG-relevant terms are  generated during RG: The coupling $\bar{g}_3\equiv\bar{\lambda}_3+i\bar{\kappa}_3$ is absent at the microscale, and it corresponds physically to a dephasing hermitian Lindblad operator, $L=\hat{\phi}^\dag(\mathbf{x})\hat{\phi}(\mathbf{x})$, while  $\bar{g}_2$  supports non-Gaussian noise generated during the course of renormalization (Fig. \ref{nongaussian} shows a sample of one-loop diagrams contributing to the flow of $\bar{g}_2$).

We summarize all the couplings introduced so far, together with their microscopic values at the ultra-violet scale, in Tab. \ref{tabmicro}. 

Before concluding this Subsection, we recapitulate how the key physical properties of the problem determine the structure of the action, $S_Q$.

\subsubsection{Retarded and Keldysh masses}
\label{effgen}

Before discussing some  diagrammatic aspects of the quantum action $S_Q$, we want to observe in this Subsection that the formation of a condensate generates not only a mass gap, $\bar{\chi}\propto \bar{\rho}_0\bar{\kappa}_c$ as usual in a conventional $\phi^4$ model, but also three additional effective Keldysh masses besides $\bar{\gamma}$. They are peculiar of the driven open nature of the system considered in this work. We refer to them in the following as retarded/advanced ($\bar{\chi}$) and Keldysh ($\bar{\gamma}$, $\bar{m}_1$, $\bar{m}_2$, $\bar{m}_3$) masses, since they all scale as a spectral gap $\sim q^2$, in the quantum scaling regime accessed with strong Markov diffusion. They serve as a convenient choice of variables to re-parametrise into dimensionless form the renormalization group flow. 

The homogeneous part of the quadratic action (obtained in Fourier space taking the limit $Q\to0$) defines the so called mass matrix through the relation
\begin{equation}\label{massmat}
M=\frac{\delta^2\Gamma_k}{\delta\bar{\Phi}^\dag(-Q)\delta\bar{\Phi}(Q)}|_{ Q=0}=
\begin{pmatrix}
0&M^A\\
M^R&M^K\\
\end{pmatrix},
\end{equation}
where we defined the vector $\bar{\Phi}=(\phi_c, \phi_c^*, \phi_q, \phi_q^*)$ in the complex basis of the fields \footnote{Notice that the practical computations are performed in real basis, following Ref. \onlinecite{Sieberer2014}.}. 
Eq. \eqref{massmat} is evaluated on the stationary state, associated to the homogeneous condensate solution ($\bar{\phi}_c=\bar{\phi}^*_c=\bar{\phi}_0$, $\bar{\phi}_q=\bar{\phi}^*_q=0$), and its matrix elements are defined as follows, 
\begin{equation}
\bar{M}^A={\bar{M}}^{R\dag}=\begin{pmatrix}
i\frac{\bar{\rho}_0\bar{\kappa}_c}{2}&0\\
0&-i\frac{\bar{\rho}_0\bar{\kappa}_c}{2}\\
\end{pmatrix},
\end{equation}


\begin{equation}
M^K=\begin{pmatrix}
i\left(\bar{\gamma}+\frac{\bar{\rho}_0\bar{g}_1}{2}\right)&\frac{\bar{g}_3^*\bar{\rho}_0}{2}\\
-\frac{\bar{g}_3\bar{\rho}_0}{2}&i\left(\bar{\gamma}+\frac{\bar{\rho}_0\bar{g}_1}{2}\right)
\end{pmatrix}.
\end{equation}
 


As anticipated,  the sole term coming from the retarded mass sector, $M^R$, is the spectral gap, $\bar{\chi}=\frac{\bar{\rho}_0\bar{\kappa}_c}{2}$.

Analogously, from the Keldysh mass sector, we extract three effective masses, $\bar{m}_1^K\sim{\bar{\rho}_0\bar{g}_1}$, $\bar{m}_2^K\sim{\bar{\rho}_0\bar{\lambda}_3}$, $\bar{m}_3^K\sim{\bar{\rho}_0\bar{\kappa}_3}$, which result from the interplay of a non-vanishing condensate ($\bar{\rho}_0\neq0$) and non-linear noise terms in the anti-hermitian part, $S_a$, of the interaction action. 
The three masses, $\bar{m}_1^K$, $\bar{m}_2^K$, $\bar{m}_3^K$, depend on the condensate density as $\bar{\chi}$, they have the same canonical scaling dimension, $[\bar{\chi}]=2$, and their diagrammatic corrections present all an ultra-violet divergence with the same physical origin in the tadpole renormalising $\bar{\rho}_0$ (see next Subsection). 
According to their definition,
$\bar{m}_1^K$, $\bar{m}_2^K$, $\bar{m}_3^K$ are tuned to zero, in the same way in which the spectral gap, $\bar{\chi}$, is fine tuned to zero, when $\bar{\rho}_0\to0$.

\subsubsection{Diagrammatics of the quantum dynamical field theory}
\label{diagrams}

We  now compare the diagrammatic structure of the action of the quantum dynamical field theory, $S_Q$, with the semiclassical model for the  condensation in three dimensions of a Bose gas coupled to  generic Markovian noise \cite{Sieberer13} (where $\bar{\gamma}_d=0$ and $\bar{u}_q=\bar{g}_1=\bar{g}_2=\bar{g}_3=0$ from the outset, on the basis of RG irrelevance). 

We compare diagrammatically the dynamical semi-classical field theory and the quantum one for two reasons. First, it grounds the different   stability properties of the classical and quantum fixed points, on the basis of the number of UV divergent diagrams occurring in the two theories. Second, it provides a qualitative understanding of the diverse technical complexity pertinent to the two problems.\\

We start discussing corrections to the spectral (or retarded), $\bar{\chi}$, and Keldysh masses, $\bar{\gamma}$.
At one loop level (and in the absence of a background field, $\bar{\phi}_0=0$), such corrections are tadpoles, supported by the classical coupling, $\bar{g}_c$, for the former, and by the multiplicative noise, $\bar{g}_1$, for the latter, as shown in the upper panel of Fig. \ref{tadpole}.  Notice that in the semi-classical problem, instead, the leading  correction to $\bar{\gamma}$ at one loop, is a  diagram $\propto \bar{g}_c^2$, with two external insertions of the condensate field, $\bar{\phi}_c=\bar{\phi}_c^*\sim\sqrt{\bar{\rho}_0}$ (see also Ref. \cite{Sieberer2014} and bottom line diagram in Fig. \ref{tadpole}). Indeed, in order to have an RG flow of the homogeneous Markovian noise, two-loop effects are necessary, and they are effectively captured within our background field method by one-loop diagrams.

Such diversity reflects in a different number of ultraviolet divergent diagrams for the two theories, which is at the root of different stability properties of the classical and quantum non-equilibrium fixed points (see Section \ref{NEQFPWF}). For the following analysis, it is sufficient to consider diagrams at the leading order in perturbation theory. 

\begin{figure}[t!]\centering
\includegraphics[width=9.2cm]
{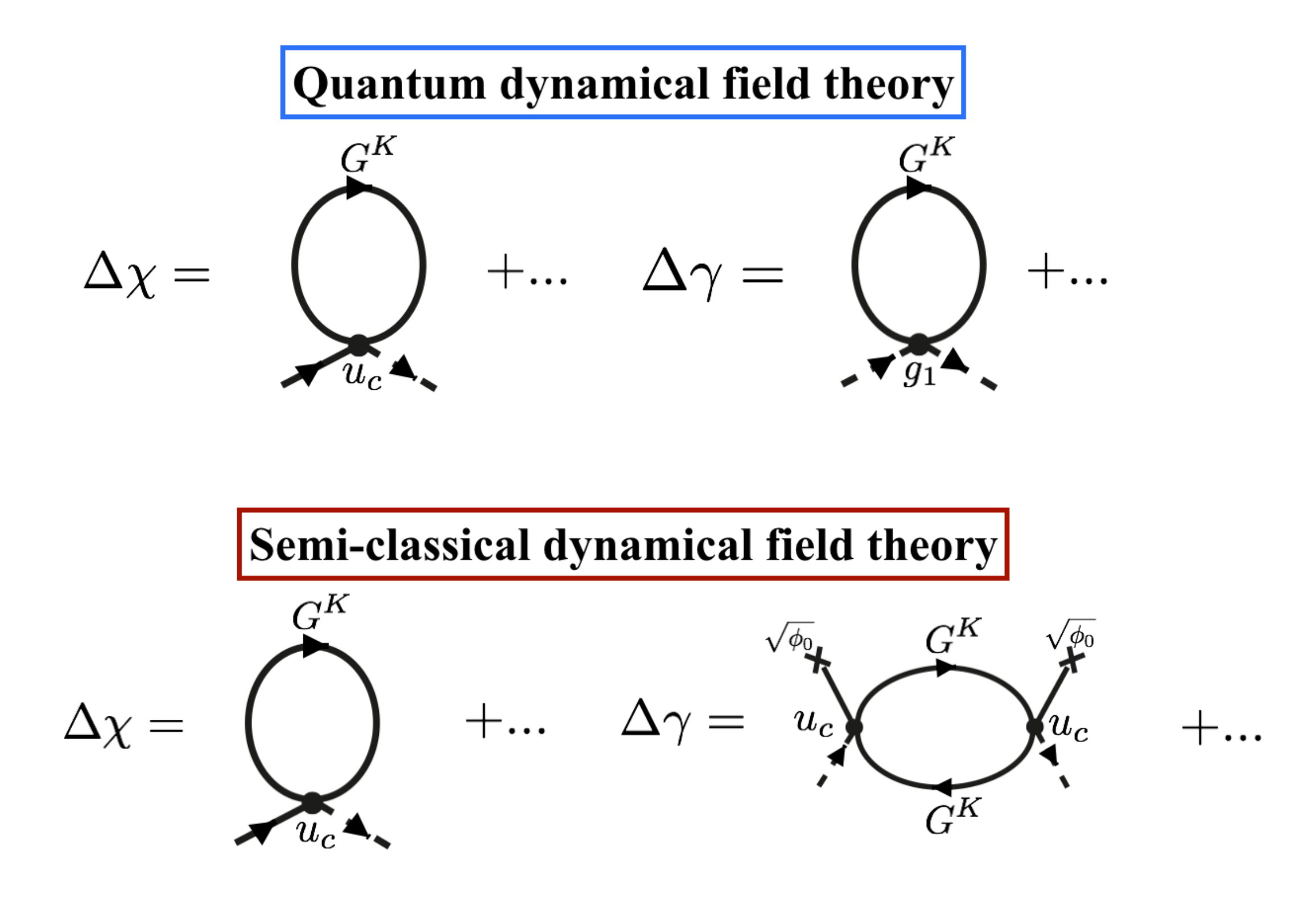}
 \label{tadpole}\caption{(Color online)
One-loop corrections (tadpoles) to the spectral, $\bar{\chi}$, and Keldysh mass, $\bar{\gamma}$. A Keldysh Green's function, $G^K$, occurs as  internal line in  tadpoles. In the non-equilibrium semi-classical model, the diagram contributing to $\Delta\bar{\gamma}$ is, instead, a two-loop diagram where one  internal Keldysh line has been cut and the two resulting external lines couple to the expectation values of the fields in the ordered phase, which is proportional to $\sqrt{\bar{\phi}_0}$. Notice that the leading correction to the condensate density, ${\bar{\rho}}_0$, is also determined by the tadpole occurring in $\Delta\bar{\chi}$. Depending whether the truncation for $\Gamma_k$ is centred in the ordered ($\bar{\rho}_0\neq0$) or disordered phase ($\bar{\rho}_0=0$) \cite{berges02:_nonper}, one has to renormalize directly $\bar{\chi}$ or the condensate density $\bar{\rho}_0$, through which the retarded mass is defined ($\bar{\chi}\sim\bar{\rho}_0\bar{\kappa}_c$) (see related discussion in Sec. \ref{microscop} and \ref{effgen}).} \label{tadpole}
\end{figure}

\begin{figure}[h!]\centering
\includegraphics[width=9.2cm]
{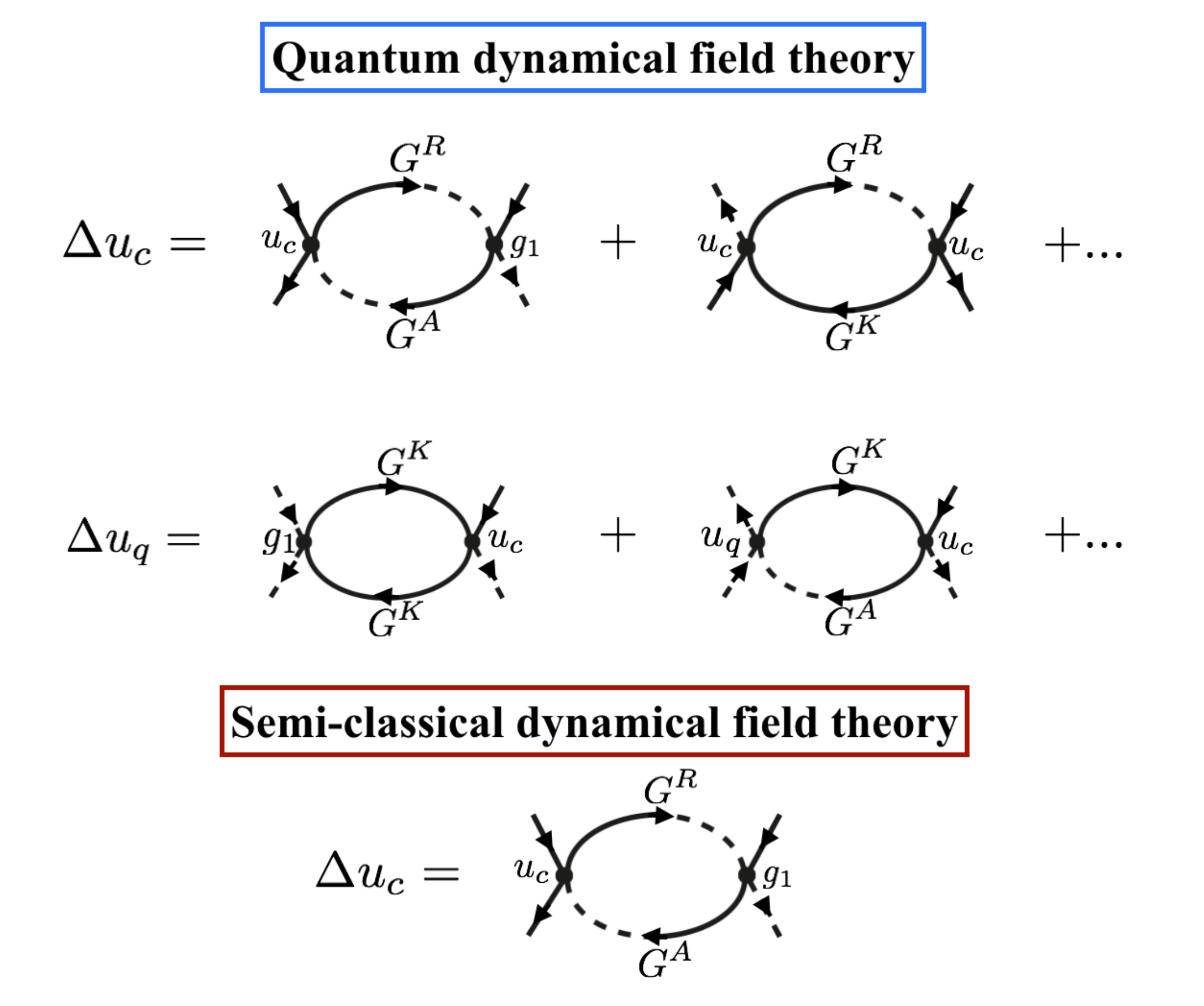}
 \caption{(Color online) 
Sample of diagrams occurring as one-loop corrections to the quartic couplings in the quantum dynamical field theory. We distinguish three classes of one-loop diagrams, containing the combinations $\bar{G}^{R/A}\bar{G}^{A/R}$, $\bar{G}^{R/A}\bar{G}^{K}$, and $\bar{G}^{K}\bar{G}^{K}$. We report  a subset of the corrections to quartic couplings (specifically, the ones related to the coherent couplings of the quartic action). On the contrary, there is  a single one-loop diagram, proportional to $\bar{g}^2_c$, in the diagrammatics of the semi-classical action, since $\bar{\phi}^{*2}_c\bar{\phi}_c\bar{\phi}_q$ is the only RG relevant operator.}
\label{oneloop}
\end{figure}

In $S_Q$  each of the two tadpoles correcting $\bar{\chi}$ and $\bar{\gamma}$ contribute with a linear ultraviolet divergence, while in the semi-classical model the sole ultraviolet divergence comes from the tadpole correcting the retarded mass, since the leading correction to $\bar{\gamma}$ (second diagram in bottom line of Fig. \ref{tadpole}), is ultraviolet convergent. 
This can be gleaned noticing that the diffusion noise produces a dimensional enhancement in the momentum integration measure occurring in diagrams: since $\bar{P}^K\sim\bar{\gamma}_dq^2$, for the tadpole correcting $\bar{\chi}$, we have,
\be
\Delta\bar{\chi}\sim \int d\omega  dq  \frac{\bar{P}^K}{\bar{P}^R\bar{P}^A}\sim  \int d\omega  dq  \frac{\bar{\gamma}_dq^2}{\bar{P}^R\bar{P}^A},
\ee 
and through comparison with the analogous diagram in the semi-classical action (where $\bar{P}^{K}\sim\bar{\gamma}$),
\be\begin{split}
\Delta\bar{\chi}&\sim \int d\omega  d^3q  \frac{\bar{P}^K}{\bar{P}^R\bar{P}^A}\sim  \int d\omega  d^3q  \frac{\bar{\gamma}}{\bar{P}^R\bar{P}^A}\\
&\sim  \int d\omega  dq  \frac{\bar{\gamma}q^2}{\bar{P}^R\bar{P}^A},
\end{split}\ee
we observe that the momentum contribution coming from the diffusion noise in loops of the one-dimensional quantum problem, is taken by the usual thee-dimensional momentum integration measure in the semi-classical action. Then, in $S_Q$, both $\Delta\bar{\chi}$ and $\Delta\bar{\gamma}$ diverge linearly for  ultraviolet momenta, while in the semi-classical problem the sole ultraviolet divergence comes from $\Delta\bar{\chi}$.
The number of ultraviolet divergences corresponds to counter-terms which should be added to the action in order to renormalise the theory. Physically, these counter-terms correspond to external parameters that has to be fine tuned in order to reach the critical regime \cite{tauberbook}. Accordingly, it is expected that the fixed point of the semi-classical theory is more stable than the quantum one, because it has a higher number of ultraviolet divergent tadpoles.

Notice that in the ordered phase ($\bar{\rho}_0\neq0$), the retarded mass, $\bar{\chi}$, is traded for the condensate density, $\bar{\rho}_0$, since $\bar{\chi}\sim\bar{\rho}_0\bar{\kappa}_c$, as previously discussed in  Sec. \ref{microscop} and \ref{effgen}. However, the  above analysis still applies, since  the tadpole correcting $\bar{\chi}$ renormalises as well the condensate density $\bar{\rho}_0$ (see Refs. \cite{berges02:_nonper, Sieberer2014}).

We now compare a class of one loop diagrams (Fig. \ref{oneloop}) contributing to corrections of  quartic couplings. In a  non-equilibrium problem loop corrections can display a rich diversity compared to equilibrium models, since they can be realised with several combinations of $\bar{G}^{R/A/K}$,  and in the absence of a thermal fluctuation-dissipation theorem, the Keldysh Green's function  is not fixed by  $\bar{G}^{R/A}$ and the thermal distribution function. In the semi-classical problem, such a complexity is ruled out from renormalization group arguments, since non-linear operators in the quantum fields are irrelevant ($\bar{u}_q=\bar{g}_1=\bar{g}_2=\bar{g}_3=0$), and this restricts the class of diagrams to the only one portrayed in Fig. \ref{oneloop}. On the other hand, the  diversity of relevant couplings in the quantum problem supports a richer morphology of diagrams (see upper part of Fig. \ref{oneloop}). The effect of dimensional enhancement of loop corrections through Markovian diffusion, is also present here, and all the corrections to quartic couplings are ultraviolet convergent. However, Fig. \ref{oneloop} elucidates not only the different complexity of the two problems, but will also help  to clarify the absence of a mapping between the critical behaviour of the quantum one-dimensional  and the classical three-dimensional Bose gases (see Section \ref{absmap}).

\subsection{FRG flow equations }
\label{flowsection}

Loop corrections to couplings can be extracted from Wetterich's equation, Eq. \eqref{Wetterich},  following the technical steps reported in Appendix \ref{app:C}. We denote the resulting beta functions for interaction couplings, ${\mathbf{g}}=\{\bar{\gamma},\bar{u}_c,\bar{u}_q,\bar{g}_1,\bar{g}_2,\bar{g}_3\}$, in compact notation, as
\be\label{betaFRG}\partial_t{\mathbf{g}}=\beta({\mathbf{g}}).\ee
Here we introduce rescaled variables using the quantum canonical scaling of Sec. \ref{QCanonical}, in order to reduce the number of flow equations \eqref{betaFRG}. We define a number of anomalous dimensions which serve also to isolate the number of independent critical exponents in the universality class. Finally, rewriting the beta functions into a set of flow equations for rescaled variables, facilitates the search for a fixed point of the scaling solution.

We  introduce rescaled fields 
\be
\phi_c=\bar{\phi}_c, \quad \phi_q=Z\bar{\phi}_q,
\ee
reabsorbing the $Z$ factors in the renormalized kinetic coefficients and  interaction couplings, accordingly defined as  follows
\be\label{zetaresc} \begin{split} K&=\bar{K}/Z,\quad \gamma_d= \bar{\gamma}_d/|Z|^2, \quad \gamma=\bar{\gamma}/|Z|^2, \\
u_c&= \bar{u}_c/Z,\quad u_q= \bar{u}_q/(Z^*|Z|^2),  \quad g_1=\bar{g}_1/|Z|^2,\\ 
 g_2&= \bar{g_2}/|Z|^4, \quad g_3= \bar{g}_3/Z^2.
\end{split}\ee 
The beta functions, then, read
\begin{equation}
  \label{beta}
  \begin{split}    
\partial_t \lambda_c & =\beta_{\lambda_c}=\eta_{Z_R} \lambda_c - \eta_{Z_I} \kappa_c +\Delta \lambda_c, \\ 
\partial_t \kappa_c & =\beta_{\kappa_c}= \eta_{Z_I} \lambda_c + \eta_{Z_R} \kappa_c +\Delta \kappa_c, \\ 
\partial_t \lambda_q & =\beta_{\lambda_q}=3\eta_{Z_R} \lambda_q +  \eta_{Z_I} \kappa_q +\Delta \lambda_q, \\ 
\partial_t \kappa_q & =\beta_{\kappa_q}=3\eta_{Z_R} \kappa_q -  \eta_{Z_I} \lambda_q +\Delta  \kappa_q, \\ 
\partial_t g_1 & =\beta_{g_1}=2\eta_{Z_R} g_1 +\Delta g_1, \\ 
\partial_t g_2 & =\beta_{g_2}=4\eta_{Z_R} g_2 +\Delta  g_2, \\ 
\partial_t \lambda_3 & =\beta_{\lambda_3}=2\eta_{Z_R} \lambda_3 -  2\eta_{Z_I} \kappa_3 +\Delta  \lambda_3, \\ 
\partial_t \kappa_3 & =\beta_{\kappa_3}=2\eta_{Z_R} \kappa_3 + 2\eta_{Z_I} \lambda_3 +\Delta  \kappa_3, \\ 
\partial_t \gamma & =\beta_{\gamma}=2\eta_{Z_R} \gamma +\Delta \gamma, \\ \end{split}
\end{equation}
where we introduced the real and imaginary parts of the anomalous dimension of the wave function coefficient, $\eta_Z\equiv\eta_{Z_R}+i\eta_{Z_I}\equiv-\partial_tZ/Z$ (a consequence of rescaling by $Z$ in Eqs. \eqref{zetaresc}). Details about the extraction of loop corrections to couplings, $\Delta \lambda_c$,...,$\Delta\gamma$, are discussed in Appendix \ref{app:B}.

For future convenience, we now trade the real parts of $u_c$ and $u_q$ by the ratios, $r_U=\frac{\lambda_c}{\kappa_c}$, $r_{U}^Q=\frac{\lambda_q}{\kappa_q}$, and we accordingly substitute the beta functions of the coherent couplings $\lambda_c$ and $\lambda_q$ with the beta functions of their associated ratios 
\begin{equation}
\label{ratios}
\begin{split}
   \partial_t r_{U} & = \beta_{r_{U}} =
  \frac{1}{\kappa_c} \left( \beta_{\lambda_c} - r_{U} \beta_{\kappa_c} \right),\\ 
  \partial_t {r_{U}^Q} & = \beta_{{r^Q_{U}}} = \frac{1}{\kappa_q}
  ( \beta_{\lambda_q} - {r^Q_{U}} \beta_{\kappa_q} ).\\
  \end{split}
  \end{equation}
Similarly, we define  the ratio of coherent ($K_R$) over dissipative ($K_I$) coefficients of the Laplacian in the retarded/advanced sector, $r=K_R/K_I$,    and its beta function as

\begin{equation}\label{erreK}
 \partial_t r  = \beta_{r} =r({\eta}_{K_I}-\eta_{K_R}), 
  \end{equation}
where we introduced the anomalous dimensions of the renormalized kinetic coefficient $K$, 
\begin{equation}
\begin{split}
\eta_{K_R}=&\bar{\eta}_{K_R}-\eta_{Z_R}+\frac{\eta_{Z_I}}{r},\\
\eta_{K_I}=&\bar{\eta}_{K_I}-\eta_{Z_R}-\eta_{Z_I}r.\\
\end{split}
\end{equation}
The bare anomalous dimensions, $\eta_{Z_R},\eta_{Z_I},\bar{\eta}_{K_R},\bar{\eta}_{K_I}$, can be extracted from  Wetterich's equation, Eq. \eqref{Wetterich}, following the procedure discussed in Appendix \ref{app:C}.\\

In order to find the quantum scaling solution of the effective action, we additionally introduce dimensionless variables according to the canonical power counting, introduced in Sec. \ref{QCanonical}.

Following the scaling analysis of Sec. \ref{QCanonical}, we characterise dimensionless retarded and Keldysh masses, 
 
\be\label{massdless} \tilde{\chi}=\frac{2\kappa_c\rho_0}{K_Ik^2}, \quad \tilde{\gamma} = \frac{\gamma}{k^2\gamma_d },\ee
likewise the three dimensionless effective Keldysh masses (see discussion in Section \ref{effgen}) 
 
 \be\label{masseffdless}\tilde{m}^1_{\textsl{K}}=\frac{2g_1\rho_0}{\gamma_dk^2},\quad \tilde{m}^2_{\textsl{K}}=\frac{2\lambda_3\rho_0}{\gamma_d k^2},\quad \tilde{m}^3_{\textsl{K}}=\frac{2\kappa_3\rho_0}{\gamma_d k^2},\ee  
and the dissipative rescaled couplings 
 \be\label{coupldless}
 \tilde{\kappa}_c =\frac{1}{4\pi} \frac{ \gamma_d
    \kappa_c}{K_I^2 k}, \quad \tilde{\kappa}_q =\frac{1}{4\pi} \frac{
    \kappa_q}{ \gamma_d k}, \quad \tilde{g}_2 =\frac{1}{4\pi} \frac{ K_I
    g_2}{ \gamma_d^2k}.
    \ee
    
In this way, the flow equations \eqref{massdless}, \eqref{masseffdless}, \eqref{coupldless}, the flow equations are  brought into dimensionless form, $\partial_t\tilde{\mathbf{g}}=\beta(\tilde{\mathbf{g}})$, and they acquire a contribution from the running of the anomalous dimensions, $\eta_a=-k\partial_k \ln a$, with $a=K_R,K_I,\gamma_d$, which are algebraic functions of the rescaled variables $\tilde{\mathbf{g}}=(\tilde{\chi},\tilde{\gamma},\tilde{\kappa}_c,\tilde{\kappa}_q,\tilde{g}_2,\tilde{m}^1_{\textsl{K}},\tilde{m}^2_{\textsl{K}},\tilde{m}^3_{\textsl{K}})$:

\begin{equation}
  \label{mass}
\begin{split}
  \partial_t \tilde{\chi} &= \beta_{\tilde{\chi}}=- \left( 2 - \eta_{K_I} \right) \tilde{\chi} + \frac{\tilde{\chi}}{\kappa_c}
  \beta_{\kappa_c} + \frac{2 \kappa_c}{k^2 K_{I}} \beta_{\rho_0},\\
 \partial_t {\tilde{m}^1_K} &= \beta_{\tilde{m}^1_K}=- \left( 2 -\eta_{\gamma_d} \right) \tilde{m}^1_K +\frac{2}{\gamma_d k^2}(\rho_0\beta_{g_1}+g_1\beta_{\rho_0}),\\
  \partial_t {\tilde{m}^2_K} &=\beta_{\tilde{m}^2_K}=- \left( 2 -\eta_{\gamma_d} \right) \tilde{m}^2_K +\frac{2}{\gamma_d k^2}(\rho_0\beta_{\lambda_3}+\lambda_3\beta_{\rho_0}),\\
 \partial_t {\tilde{m}^3_K} &=\beta_{\tilde{m}^3_K}=- \left( 2 -\eta_{\gamma_d} \right) \tilde{m}^3_K +\frac{2}{\gamma_d k^2}(\rho_0\beta_{\kappa_3}+\kappa_3\beta_{\rho_0}),\\
  \partial_t \tilde{\gamma} &=\beta_{\tilde{\gamma}}=- \left( 2 -\eta_{\gamma_d} \right) \tilde{\gamma} +\frac{1}{\gamma_d k^2}\beta_\gamma,\\
   \partial_t \tilde{\kappa}_c &= \beta_{\tilde{\kappa}_c} = - \left( 1 - 2 \eta_{K_I}
    + \eta_{\gamma_d} \right) \tilde{\kappa}_c + \frac{1}{4\pi}\frac{  \gamma_d}{ k {K^2_I}}
  \beta_{\kappa_c}, \\
   \partial_t \tilde{\kappa}_q&= \beta_{\tilde{\kappa}_q}= -\left( 1- \eta_{\gamma_d} \right) \tilde{\kappa}_q +  \frac{1}{4\pi}\frac{1}{\gamma_d k} \beta_{\kappa_q},\\
   \partial_t \tilde{g}_2 &= \beta_{\tilde{g}_2}=- \left( 1 +  \eta_{K_I}
    -2 \eta_{\gamma_d} \right) \tilde{g}_2 + \frac{1}{4\pi} \frac{ K_I }{ \gamma_d^2k}
  \beta_{{g}_2}.
\end{split}
\end{equation}
Notice that the apparently dimensionful combinations in the right hand side of Eq. \eqref{mass}, can be easily expressed in terms of the dimensionless variables, $\tilde{\mathbf{g}}$, and ratios, $r, r_U, r^Q_U$.
 
Loop corrections to the condensate density determine the beta function $\beta_{\rho_0}$ \cite{berges02:_nonper, sieberer13:_dynam_critic_phenom_driven_dissip_system}.  In Eqs. \eqref{mass} we also defined  the anomalous dimension of the diffusion noise coupling, $\gamma_d$, (see App. \ref{app:C} for further details)
\begin{gather}
\eta_{\gamma_d}={\bar{\eta}}_{\gamma_d}-2\eta_{Z_R}.
\end{gather}

Setting to zero the right hand sides of the dimensionless renormalized flow equations, Eqs.~\eqref{mass}, we find a Wilson-Fisher fixed point (FP) characterised by \be\label{FPquantum}
\begin{split}
r^*=&1.55, \quad r^*_U=0.397, \quad {r^Q_U}^*=-0.248, \quad \tilde{\kappa}_c^*=0.146,\\
\tilde{\kappa}_q^*=&-0.028,\quad \tilde{g}^*_2=0.0047,\quad  \tilde{\chi}^*=0.900, \quad \tilde{\gamma}^*=0.023, \\
{\tilde{m}^{1*}}_{\textsl{K}}=&0.22, \quad{\tilde{m}^{2*}}_{\textsl{K}}=-0.0484, 
\quad{\tilde{m}^{3*}}_{\textsl{K}}= 0.803.\\
\end{split}
\ee

\section{Critical properties of the driven-diffusive Bose gas} 
\label{NEQFPWF}

We will now use as a benchmark for the salient physical features of this quantum FP in one dimension (with noise level $P^K= \gamma+2\gamma_dq^2$), its semiclassical driven Markovian counterpart in $d+z=3$ dimensions, which results from a similar FRG analysis \cite{Sieberer13} of the driven open Bose gas subject to flat Markovian noise ($P^K= \gamma$).  This comparison is utile to highlight the absence of a mapping between  the quantum and classical  driven-dissipative condensation transitions. 

We can already notice that the key new property of the quantum FP is its mixed nature with coexistent coherent and dissipative processes, manifest in non-vanishing ratios (cf Eqs. \eqref{FPquantum}). This is in clear contrast with the Wilson-Fisher fixed point of the driven-dissipative semi-classical condensation  in three  dimensions \cite{Sieberer13}, where $r^*=r^*_U=0$, realising a purely dissipative situation  (see also Fig. \ref{Fig2} for a pictorical comparison).

\subsection{Stability matrix}

We now  discuss in more detail the structure of the stability matrix of the quantum fixed point \eqref{FPquantum}. This will also allow us to discuss  in the next subsection a key  property of the novel critical regime -- absence of decoherence at long wavelengths. 

In order to extract $S$, we expand the dimensionless beta functions, $\partial_t{\tilde{\mathbf{g}}}=\mathbf{\beta}({\tilde{\mathbf{g}}})$, at first order in  small displacements, $\delta {\tilde{\mathbf{g}}}= {\tilde{\mathbf{g}}}-{\tilde{\mathbf{g}}}^*$, around the  FP values of the couplings, Eq. \eqref{FPquantum}. In this way we   introduce the  stability matrix, $S$, 
\be
\partial_t \delta {\tilde{\mathbf{g}}}=S\cdot\delta\tilde{\mathbf{g}}.
\ee
$S$ does not have a block structure: 
this implies that all the masses present in $S_Q$ (recall the discussion in Section \ref{effgen}) scale in the same way at the FP with the largest negative eigenvalue, $-\lambda_S$, of $S$, dominating the asymptotic infrared properties. Specifically, $\chi,\gamma, m^1_K, m^2_K, m^3_K\sim k^{\lambda_S}$, for $k\to0$, with $\lambda_S=2.47$. From $\lambda_S$, it
also easy to extract the correlation length critical exponent, $\nu=1/\lambda_S=0.41$, which does not coincide with the $O(2)$ equilibrium value, $\nu\simeq 0.70$, extracted from FRG, see also \onlinecite{berges02:_nonper} ($\nu$ is the same also for the driven-dissipative condensation in three dimensions). This constitutes a further hint (in addition to the diagrammatic complexity glimpsed in Sec. \ref{diagrams}) that the universality class of driven Markovian quantum criticality does not fall into an equilibrium universality class, or in the one of its higher dimensional counterpart. Each independent block of the stability matrix determines an independent exponent: besides $\nu$ associated to $S$, anomalous dimensions at the quantum FP  are  evaluated substituting in the expressions for $\eta_a$ ($a=Z_R,Z_I,K_R,K_I,\gamma_d$), the FP values, Eq. \eqref{FPquantum} (see for a summary Tab. \ref {tabella}).

The stability matrix of the novel quantum fixed point possesses as many repulsive directions (negative eigenvalues), as many   masses are in the model (see Section \ref{effgen}), and each one of them carries an ultra-violet diagrammatic divergence, as expected from the RG theory for critical phenomena \cite{Amit/Martin-Mayor, Zinn-Justin, Cardy}. The ultra-violet divergences in the RG flow of $\chi$, $m^1_K$, $m^2_K$, $m^3_K$ share the same diagrammatic and physical origin, since they all emanate from the ultra-violet divergence of the tadpole renormalizing $\beta_{\rho_0}$, which occurs in their respective flow equations, see Eqs. \eqref{mass}. For the Markovian noise, $\gamma$, the ultra-violet divergence comes instead from the tadpole contributing to $\beta_\gamma$ (recall Fig. \ref{tadpole} and related discussion).

\subsection{Absence of asymptotic decoherence}
\label{decohsec}

\begin{table}[]
\centering
\begin{tabular}{|l|l|l|l|l|l|l|l|l|}
\hline
Critic. Exps. &$\nu$ & $\eta_{K_R}$ & $ \eta_{K_I} $ &  $\eta_{Z_R} $  & $\eta_{Z_I} $  &  $\eta_{\gamma_d}$ & $\eta_{\gamma}$   \\ \hline
DD Quantum &0.405  &-0.025  &  -0.025&0.08 & 0.03 &  -0.26 & $\times$    \\ \hline
DD SC  &0.72  &-0.22  &-0.12    &  0.16& 0 & $\times$  & -0.16  \\ \hline
\end{tabular}
\caption{Comparison between the critical exponents of an open Bose gas in the presence of flat Markovian noise, described by a driven-dissipative semi-classical  action, and driven-diffusive open Bose gas, discussed here. Within the semi-classical scaling, $\gamma\sim k^0$ at the canonical level, and the Markovian noise can acquire an anomalous dimension, $\eta_{\gamma}$; this role is taken by $\eta_{\gamma_d}$ in the quantum case.  Since the two sets of critical exponents are not identical, the universality class of the one-dimensional  quantum driven-dissipative Bose gas and of its three dimensional  classical counterpart do not coincide.}\label{tabella}
\end{table}

\begin{figure}[t!]\centering
\includegraphics[width=8.3cm] {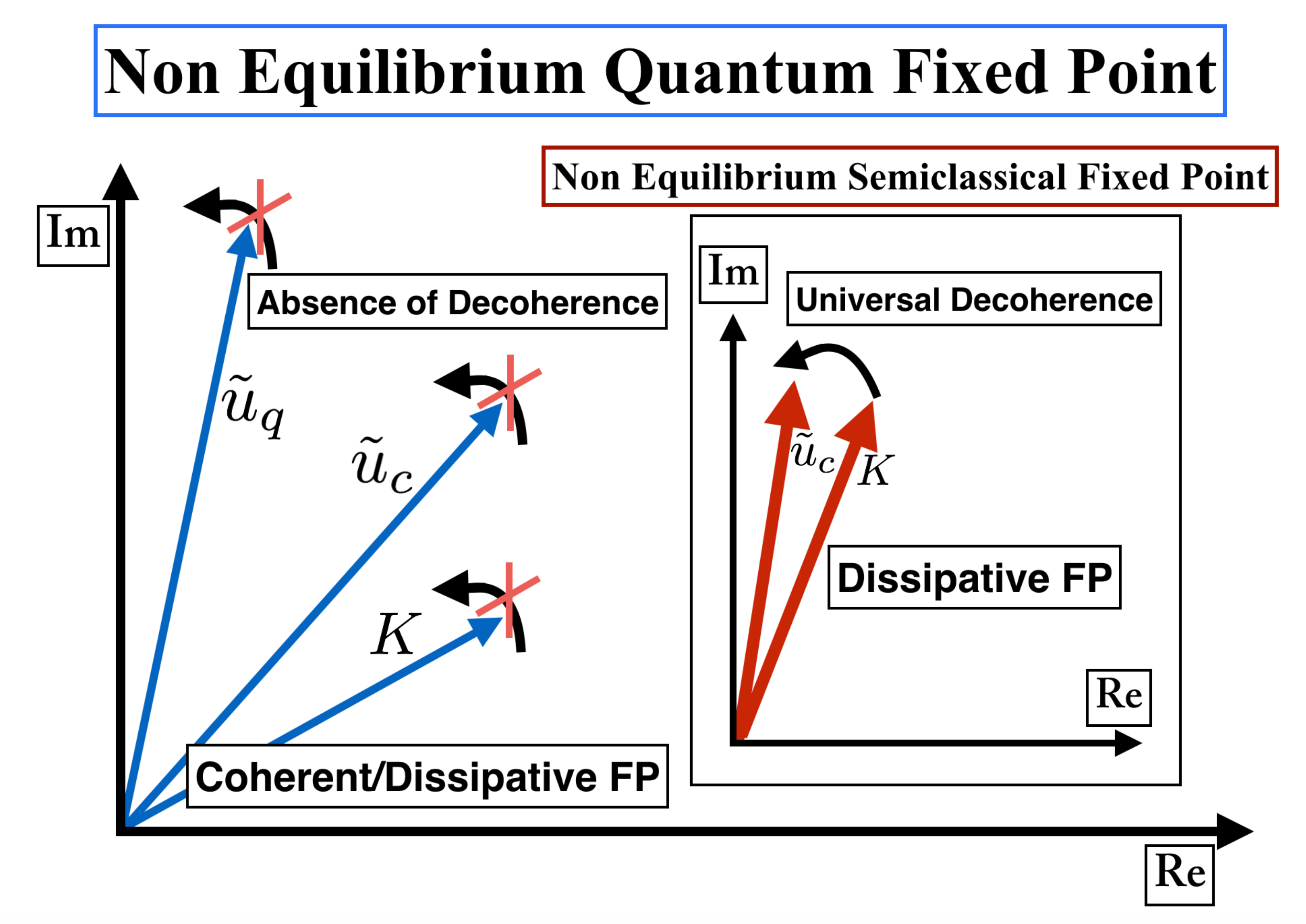}
 \caption{(Color online) Comparison between the fixed points of the non-equilibrium quantum action and of the semi-classical action of driven-dissipative Bose gases.  The fixed point of the former displays simultaneous presence of finite dissipative and coherent couplings (absence of decoherence).  The fixed point of the latter is instead entirely dissipative (decoherence at long wavelengths), and at infrared momenta  the couplings flow towards the imaginary axis.}
\label{Fig2}
\end{figure}

Tab. \ref {tabella} shows that there is an
 exponent degeneracy linking the anomalous dimensions of the retarded (real and imaginary) kinetic coefficients, $\eta_{K_R}=\eta_{K_I}=-0.025$. Such circumstance allows for a finite ratio of coherent propagation ($K_R$) versus diffusion ($K_I$), $r\sim k^{-\eta_{K_R}+\eta_{K_I}}$ (real and imaginary parts of the coefficient of the Laplacian in the retarded/advanced sector), which is   consistent with the finding of finite ratios at the FP, Eq. \eqref{FPquantum}. In particular, it indicates  absence of decoherence at long distances, i.e. persistence of quantum mechanical facets at criticality (see Fig. \ref{Fig2}), since neither dissipation completely dominates, leading to a FP value of $r^*\to0$, nor quantum coherent effects entirely dictate the infrared physics, $r^*\to\infty$ (as it would be the case for a zero temperature quantum phase transition). 

Survival of quantum coherence at distances smaller than the Markov length, ${\Lambda_M}^{-1}$,  is a similarity  between the FP of quantum driven-dissipative condensation and equilibrium quantum critical points at finite temperature , where thermal fluctuations destroy coherent effects only at distances larger than the de Broglie length \cite{sondhi, Sachdev}.


Besides the fixed point \eqref{FPquantum},  the RG flow of the quantum dynamical field theory admits also a \emph{purely dissipative} fixed point ($r^*=r^*_U=r^Q_U=0$), which, nevertheless, can never be reached, as we comment below.  On the other hand the fixed point of the driven open gas in three dimensions admits as a stable fixed point only the \emph{purely dissipative} one (see Fig. \ref{Fig2} for a comparison with the fixed point \eqref{FPquantum}). We are now going to elucidate this important difference, since  it further consolidates  the absence of decoherence as a key property of the  quantum non-equilibrium critical regime.

Let us first summarise some properties of the stability matrix of a dissipative FP. It has the following block-diagonal structure:
\be
S=\begin{pmatrix}D &0\\
0&R
\end{pmatrix};
\ee
$D$ is expressed in the basis spanned by purely imaginary couplings $\{\tilde{m}_R, \tilde{\kappa}_c, ...\}$, while $R$ is written in  the subspace of ratios $\{ r, r_U, ...\}$. The negative eigenvalue associated to the fine tuning of the spectral mass (and to the correlation length critical exponent, $\nu$) comes from the sub-block $D$, while the subspace $R$ should be attractive for the RG flow  to have a physically realisable purely dissipative FP 
: fine tuning of  retarded and Keldysh masses can only affect the stability properties of the $D$ sector, and then $R$ should possesses only positive eigenvalues in order to be attractive under RG flow.

The block $R$  contains also new important information on the critical properties, since it determines a novel anomalous dimension, $\eta_r$, which controls infrared decoherence close to the fixed point of  $O(2)$ models. However, its value is sensitive to the presence or absence of emergent thermal equilibration at  the fixed point. In this sense, $\eta_r$ extends to a non-equilibrium setting the set of critical exponents of the $O(2)$ equilibrium universality class   in three dimensions  \cite{Sieberer13}. In particular, such anomalous dimension  controls the universal fade out of coherent couplings,
\be
r\sim r_U\sim_{k\to0}k^{\lambda_R}\sim k^{-\eta_r}\to0,
\ee
and it is determined by the smallest positive eigenvalue of the stability matrix, $\lambda_R$ (with this convention, $\eta_r$
must be negative in order to have  an  attractive dissipative FP).
For instance, for the three-dimensional driven open Bose gas coupled to homogeneous Markovian noise, it has been found $\gamma$, $\eta_r^{SC}=-0.10$.

On the other hand, the dissipative FP ($r^*=r^*_U=r^Q_U=0$)   present in the RG flow of the one dimensional driven-diffusive open  Bose gas, exhibits  $\eta^{Q}_r=0.83$, which indicates that such purely dissipative FP can never be never reached at infrared scales by the RG quantum flow, since it is  repulsive in a way that cannot be fixed by the fine  tuning of available physical masses, since it can only affect instabilities in the sector $D$ of $S$.

\subsubsection{Infrared dispersion relation of critical modes}
\label{modesec}

A physical consequence of an infrared purely dissipative FP, is a complex diffusive dispersion relation of critical modes, with leading diffusive behaviour.

Rewriting in terms of dimensionless variables the effective dispersion relation of long-wavelength excitations, we find at the semiclassical three-dimensional fixed point ($r^*=r^*_U=0$), $\omega(q)\sim-iK_Iq^2$, which yields a leading diffusive infrared behaviour for the dispersion relation at the semi-classical critical point  (subleading coherent terms are also present, see for instance Section VIII.C of Ref. \onlinecite{Sieberer2013}). 

However, at the  critical point of quantum driven-dissipative condensation, Eq. \eqref{FPquantum}, we have instead 

\be\label{renomrw}
\omega(q)=K_Iq^2\Big[-i(1+\tilde{\chi}^*/2)\pm\sqrt{r^2+rr_U\tilde{\chi}^*-(\tilde{\chi}^*/2)^2}\Big],
\ee
where we can see that   real and  imaginary parts are both leading in the dispersion relation, due to the exponent degeneracy, $\eta_{KI}=\eta_{KR}$.  Notice that for the FP values in Eq. \eqref{FPquantum}, the square root in Eq. \eqref{renomrw} has a positive argument. 

From Eq. \eqref{renomrw} we can also read off  anomalous corrections to the dynamical critical exponent $z'=2-\eta_{K_I}\simeq2.025$. It indicates \emph{a posteriori} that the  truncation, Eq. \eqref{int}, i.e. the quantum dynamical field theory, and the associated quantum canonical power counting,  have been properly chosen to capture the  critical behaviour encoded in the FP, Eq. \eqref{FPquantum}.

\subsection{RG limit-cycle oscillations of $Z$} 
\label{limitsec}

Another hallmark of the persistence of coherent quantum facets at the fixed point, are renormalization group limit-cycle oscillations imprinted on the spectral density by   oscillations (in the running scale variable $t$) of the wave function renormalization coefficient  $Z$. 

We find a non-vanishing  imaginary part of $\eta_Z$ at the quantum FP, $\eta_{Z_I}=0.03$, which determines an oscillatory wave-function renormalization coefficient, $ Z\sim k^{-\eta_{Z_R}}e^{-i\eta_{Z_I}t}$, in the RG time variable, $t\propto\log k$.

Notice from the definition of the anomalous dimension of $Z$,
\be
\eta_Z\equiv\eta_{Z_R}+i\eta_{Z_I}\equiv -\frac{\partial_t(Z_R+iZ_I)}{Z_R+iZ_I},
\ee 
that  in the vicinity of a dissipative FP, where $\eta_{Z_I}=0$, these oscillations are absent and then $Z\sim k^{-\eta_{Z_R}}$, while for the FP, Eq. \eqref{FPquantum}, we have $\eta_{Z_I}\neq0$. Accordingly, we can write the real and imaginary parts of $Z=Z_R+iZ_I$, as
\begin{equation}\label{oscillations}
\begin{split}
Z_{R}(t)&=e^{-\eta_{Z_R}t}\Big[C_1\cos{\eta_{Z_I}t}+C_2\sin{\eta_{Z_I}t}\Big],\\
 Z_{I}(t)&=e^{-\eta_{Z_R}t}\Big[C_2\cos{\eta_{Z_I}t}-C_1\sin{\eta_{Z_I}t}\Big],
 \end{split}
\end{equation}
where $C_{1,2}$ are two non-universal constants.

While the absolute value of $Z$ scales identically in the two cases, $|Z|\sim k^{-\eta_{Z_R}}$, the phase of $Z$ is a $k$-dependent quantity in the vicinity of the quantum FP, $\arg Z=\arctan (Z_I/Z_R)$, contrary to what happens close to a dissipative one.

This has physical consequences on the spectral density of the system. In a neighbourhood of its peak, $\omega\simeq \operatorname{Re} \omega_q$, the spectral density has the Lorentzian form
\begin{equation}
\begin{split}
&A(Q)=-2 \operatorname{Im} \langle\bar{\phi}^*_c(Q)\bar{\phi}_q(Q)\rangle=\\
&\frac{2}{|Z|^2}\frac{(\operatorname{Re} Z)( \operatorname{Im} \omega_q)}{(\omega-\operatorname{Re}\omega_q)^2+(\operatorname{Im}\omega_q)^2},
\end{split}\end{equation}
where $\omega_q$ is the renormalized dispersion relation, Eq. \eqref{renomrw}. 
At $\omega=\operatorname{Re}\omega_q$, we extract $A(\omega=\operatorname{Re}\omega_q)=\frac{\operatorname{Re}(Z)}{|Z|^2}\frac{1}{\operatorname{Im}\omega_q}$, which is sensitive to RG oscillations of $Z$.  
Indeed, these limit-cycle oscillations occur with a huge period, $\frac{2\pi}{\eta_{Z_I}}\approx 210$ (in units of RG time, $t$) and therefore would be hard to detect, although they are a remarkable signature of deviation from equilibrium behaviour at macroscales, preventing the presence of a real wave-function renormalization $Z$. The long oscillation period, however, guarantees that $A(Q)$ stays positive: a semi-quantitative estimate for the RG evolution of  $Z(t)$, in the momentum range $k<\Lambda_G$ (where the fixed point results holds), shows that $Z(t)$ never changes sign.

In contrast,  in the equilibrium MSRJD action of model A or in the vicinity of the semi-classical fixed point of three dimensional  condensation \cite{Sieberer13}, we find $\eta_{Z_I}=0$, and these long-period oscillations are not present. We checked that for \emph{purely dissipative} fixed points of driven-dissipative Bose systems, $\eta_{Z_I}$ always vanishes, suggesting that the limit cycle behaviour of the wave-function coefficient $Z$ is linked to the absence of decoherence at long wavelengths.

Before moving further, we notice that this oscillatory behaviour is also present in the bare kinetic coefficient of the Laplacian $\bar{K}$. Indeed, computing at the quantum FP its value
\begin{equation}\begin{split}
\eta\equiv&\eta_R+i\eta_I\equiv-\frac{\partial_t\bar{K}}{\bar{K}}=\\
&\frac{1}{1+r^2}\Big[r^2{\bar{\eta}}_{K_R}+{\bar{\eta}}_{K_I}-ir({\bar{\eta}}_{K_R}-{\bar{\eta}}_{K_I})\Big],
\end{split}\ee
it is  easy to extract the scaling $\bar{K}\sim e^{-\eta_{R}t-i\eta_{I}t}$, which, remarkably,  compensates the oscillations in $Z$ such that the ratio
\be\label{kappa}
K=\frac{\bar{K}}{Z}\sim\frac{ e^{-\eta_{R}t-i\eta_{I}t}}{e^{-\eta_{Z_R}t-i\eta_{Z_I}t}}
\ee
does not exhibit any limit-cycle oscillation. This in turn also guarantees that the poles of $\bar{G}^R$ are always located in lower half plane. 
As a check of consistency, the values of renormalized anomalous dimensions $\eta_{K_R}$ and $\eta_{K_I}$, of Tab \ref{tabella}, are reproduced by Eq. \eqref{kappa}. 

Finally, we find it interesting to mention the existence of a twin FP (with $(r^*,r_U^*,r_U^{Q*})\to(-r^*,-r_U^*,-r_U^{Q*})$, and dissipative couplings unchanged), which exhibits the same critical exponents of the quantum FP discussed so far, except for an opposite value of $\eta_{Z_I}=-0.03$, which would then imply counter-phase RG limit-cycle oscillations in $Z$.

\subsection{Absence of infrared thermalization}
\label{FDRsec}

So far we discussed the persistence of quantum facets at criticality. We now discuss the absence of any emergent equilibrium behaviour at the fixed point.
 
A convenient diagnostic tool for thermal equilibrium in quantum many body systems, is the presence of a symmetry of the Keldysh functional integral, which combines quantum-mechanical time reversal and the Kubo-Martin-Schwinger condition \cite{DiehlGambx} (see for the classical case Ref. \onlinecite{aron10:_symmet_langev}). The Ward-Takahashi identities associated to this symmetry are  thermal fluctuation-dissipation relations of arbitrary order (FDR). 

While the RG flows of purely dissipative dynamical field theories fulfils  FDR  (like the Langevin critical dynamics of model A in the Hohenberg-Halperin classification) , the RG flow of the semi-classical action for driven-dissipative condensation spoils the FDR. However, the FDR is recovered at the semi-classical FP, indicating  thermalization for the low energy modes, despite the strongly driven driven nature of the system.  
This is shown, extracting the low frequency distribution function  from the generic parametrization of the Keldysh Green's function \cite{kamenevbook}
\begin{equation}\label{FDR}
G^K=G^R\sigma^zF-F\sigma^zG^A,
\end{equation}
in terms of the hermitian matrix $F$,  the distribution function.
The Pauli matrix $\sigma^z$ has been introduced to respect the symplectic structure of the bosonic Nambu space, so that the matrix elements of $G^K$ are consistent with bosonic commutation relations \cite{Buchhold2015}.

We compute from Eq. \eqref{FDR} the leading $\omega\to 0$ behaviour of the distribution function, finding $F_C(\omega,k)\sim\frac{2T_{C}}{\omega}$. The coefficient, $T_{C}\sim |Z|\gamma$,  can be interpreted as a temperature, if it is  independent of the RG scale, while it can manifest some explicit dependence from the running RG scale, $k$, in the more generic non-equilibrium case.  In the RG formalism, thermalization is then formulated as invariance under RG partitions of the system. %
 This is just a rephrasing of the detailed balance condition in the language of RG critical phenomena \cite{Sieberer2014}. 
At the semi-classical fixed point, the emergent exponent degeneracy $\eta_\gamma=-\eta_{Z_R}$ (see Ref.\cite{Sieberer13} and Tab. \ref{tabella}), indicates that $T_C$ is scale invariant. Accordingly, restoration of the FDR, and thermal behaviour for infrared modes, emerge at the semi-classical FP of three-dimensional driven-dissipative condensation.

In the same way, a statement about thermalization in the quantum problem can be formulated in terms of the anomalous dimension of the diffusion noise coupling, $\eta_{\gamma_d}$. Indeed, $\eta_{\gamma_d}$, is pivotal to assess the equilibrium or non-equilbrium character of the system at infrared scales, as $\eta_\gamma$ is for the semi-classical fixed point.

If equilibration were to ensue close to the quantum fixed point, we would expect scale-invariance of the low-frequency distribution function, $F_Q(\omega,k)$, since in an analog of a zero temperature setting (as the scaling regime realised by the diffusion Markov noise), the fluctuation-dissipation relation implies a flat distribution function in frequency and momentum, $F\propto|\omega|/\omega~ \mathbb{I}$ (where $\mathbb{I}$ is the identity Matrix in Nambu space) \cite{kamenevbook}.

 %

Before discussing the explicit matrix expression of $F$, a technical comment is in order: we choose to renormalize the fields introducing an imaginary unit, $\hat{\phi}_c=\bar{\phi}_c$ and $\hat{\phi}_q=i\frac{Z}{|Z|}\bar{\phi}_q$. This choice has been already adopted in \onlinecite{Sieberer13} to make closer contact between the purely dissipative infrared action of the  semi-classical model for non-equilibrium condensation and  the functional integral representation of model A, where the action is purely imaginary, and infrared thermalization is expected. 

We employ here the same conventions  in order to compare more closely the results for driven-dissipative classical and  quantum criticality.
Inverting equation Eq. \eqref{FDR}, we  find in terms of the rescaled variables of Sec. \ref{flowsection}, the result
\begin{equation}\label{effeFDRG}
F(\omega,q)=\begin{pmatrix}
 \frac{\bar{\gamma}_dq^2}{|Z|}\frac{\tilde{\gamma}^*_{tot}}{\omega-K_Rq^2} & -i\frac{\bar{\gamma}_d}{\bar{K}_I}\frac{Z^*}{|Z|}\frac{{\tilde{m}}^{2*}_{\textsl{K}}-i{\tilde{m}}^{3*}_{\textsl{K}}}{8(r-i+i\frac{{w}^*}{2})} \\
 i\frac{\bar{\gamma}_d}{\bar{K}_I}\frac{Z}{|Z|}\frac{{\tilde{m}}^{2*}_{\textsl{K}}+i{\tilde{m}}^{3*}_{\textsl{K}}}{8(r+i-i\frac{{w}^*}{2})} &   \frac{\bar{\gamma}_dq^2}{|Z|}\frac{\tilde{\gamma}^*_{tot}}{\omega+K_Rq^2}    \end{pmatrix}, \end{equation}
 where $\tilde{\gamma}^*_{tot}=1+\tilde{\gamma}^*/2+{\tilde{m}}^{1*}_{\textsl{K}}/8$.
As we now discuss, Eq. \eqref{effeFDRG} violates the requirements for  effective thermalization at zero temperature.
First of all, we notice that at equilibrium the off-diagonal elements are absent, while here they are supported by the effectively generated Keldysh masses, $m_2^K$ and $m_3^K$, which have a non-vanishing FP value (see Eq. \eqref{FPquantum}). In addition to that, since $\eta_{K_R}=-0.025$, the kinetic term scales faster to zero in comparison to $\omega\sim q^2$, and  we then find for the diagonal elements ($F_{1,1}$ and $F_{2,2}$) in the infrared limit, 
\be\label{effequan}
F_Q(\omega,q)\sim\frac{\bar{\gamma}_d q^2}{|Z|\omega}\tilde{\gamma}^*_{tot}=\frac{|Z|T_Q(q)}{\omega}\tilde{\gamma}^*_{tot},
\ee
 where $T_Q(q)=\gamma_dq^2$ is the $q$-dependent 'temperature' induced by the  Keldysh diffusion, $\gamma_d$. 
 The expressions of the diagonal elements are formally similar to the classical distribution function extracted in the vicinity of the semi-classical driven-dissipative fixed point, $F_C(\omega,k)\sim\frac{T_{C}}{\omega}$. The shape of Eq. \eqref{effequan}, however, already suggests the  non-equilibrium nature of the system: $T_Q(q)=\gamma_dq^2$ is not  a physical  temperature, since it depends explicitly on momentum (cf. the previous discussion on the notion of temperature in a RG language). 
 
Moreover, we notice that even if canonically the frequency can compensate  for the $q^2$ term in the numerator of Eq. \eqref{effequan} ($\omega\sim q^z$, where $z=2$), the lack of exponent degeneracy $\eta_{\gamma_d}\neq-\eta_{Z_R}$ (cf. with Tab. \ref{tabella}) indicates the absence of infrared thermalization at the quantum FP, since for an equilibrium quantum system we must have  exactly $F\propto|\omega|/\omega~ \mathbb{I}$.

\subsection{Constraints on the renormalization group flow from thermalization}
\label{constrsec}

\begin{figure}[t!]\centering
\includegraphics[width=8.3cm] {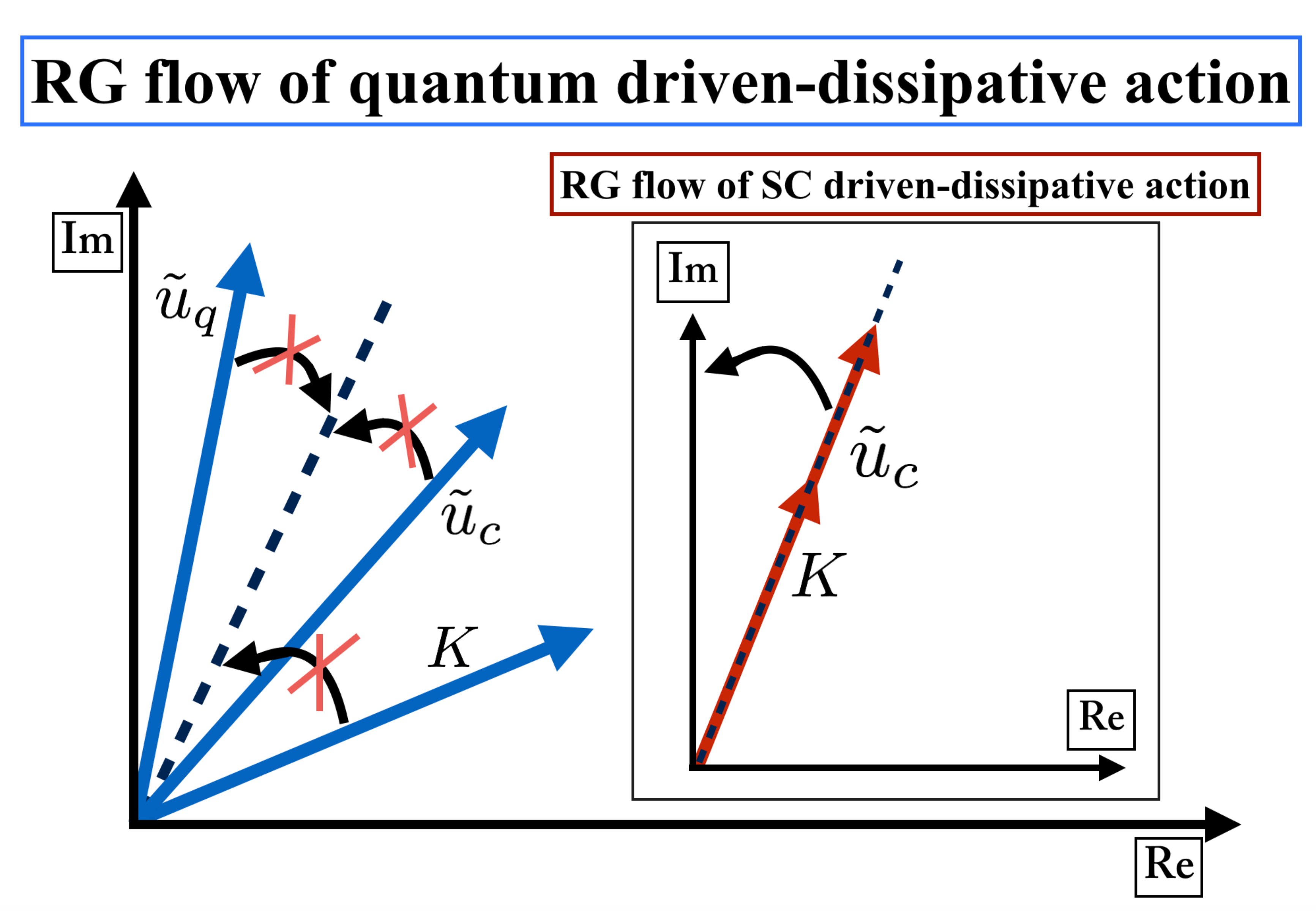}
 \caption{(Color online) A renormalization group flow with all the couplings lying  on a single ray is a sufficient condition for infrared thermalization at the fixed point, as occurs in the RG study of three dimensional semi-classical (SC) criticality, in the presence of flat Markov noise. This feature is absent in the RG flow of the quantum dynamical field theory (no common rotation towards the dashed red line in the main panel), suggesting already that the quantum fixed point cannot  possess thermal features.}
\label{align}
\end{figure}

The equilibrium symmetry, mentioned at the beginning of the previous Section, constrains the form of a driven-dissipative action equipped with coherent and dissipative couplings. For instance, it can be  implemented \emph{ab initio}  in the semi-classical model for three-dimensional driven-dissipative condensation, enforcing a common ratio between coherent and dissipative couplings, $r=r_U$. This remarkably implies a constraint on the beta functions, $\beta_r=\beta_{r_U}$. In this case, $\beta_r=\beta_{r_U}\neq0$, which  generates an 'RG force' driving the system towards  the dissipative FP, and accordingly an aligned flow of  $r$ and $r_U$ (for a sketch of the proof see Section VI of Ref. \cite{Sieberer13}, while for a graphical representation see Fig. \ref{align}). 

The presence of non-linear  noise terms in  $S_Q$, spoils, instead,   the feasibility of a flow where  the ratios evolve lined up, as well as the formation of a fixed point where all the ratios share a common value (see Fig. \ref{align}): Fixing $r=r_U$, we see that the mismatch between $\beta_r$ and $\beta_{r_U}$ is proportional to the coupling, $g_3$, which, then, is responsible for the impairing of the equilibrium symmetry during the RG flow, since it gets a non-vanishing FP value (even if $g_3$ was absent at the microscopic scale, it would be effectively generated in the course of renormalization, see discussion in Sec.~ \ref{introSQ} and the FP values in Eq. \eqref{FPquantum}).

Accordingly, the absence of emergent thermalization in the infrared flow of $S_Q$, might have  been already glimpsed from the different values acquired at the FP by the ratios $r^*\neq{r_U}^*$, ${r_U}^*\neq{r_U^{Q*}}$, $r^*\neq{r_U^{Q*}}$. 

\subsection{Lack of correspondence between the universality class of quantum and semi-classical driven Markovian criticality}
\label{absmap}

At equilibrium, quantum critical systems in $d$ dimensions (with dynamical critical exponent $z$) fall in the same universality class of their classical critical $d+z$-dimensional  counterparts. This key property constitutes a pillar in the theory of equilibrium critical phenomena, and it is intuitively understood realising that the time variable can augment the dimensionality of the system after a Wick's rotation has been performed \cite{sondhi, Sachdev}. For instance, the static critical exponents of the $d$ dimensional quantum Ising model ($z=1$) and of the $d+1$-dimensional classical one, are the same. 
A more exhaustive review for equilibrium critical systems can be found in Refs. \onlinecite{sondhi, Sachdev, vojta}.


The absence of infrared restoration of the equilibrium symmetry (thermalization) and of  the thermal fluctuation-dissipation relation, constitutes already a  
strong evidence that the quantum  universality class discussed in this work ($d=1$) cannot be related to its  semi-classical driven Markovian counterpart in $d+z=3$ dimensions, or to any equilibrium fixed point: the lack of a symmetry is oftentimes at the root of the mismatch of   two universality classes\cite{Amit/Martin-Mayor, tauberbook}. 

In the following, we provide a diagrammatic argument to clarify which features of the quantum dynamical field theory   are peculiar, in order to violate the correspondence with its semi-classical critical counterpart. 
In this argument,  quartic \emph{quantum} couplings which are  non-vanishing at the quantum fixed point, play a pivotal role. 

In loop corrections containing a single Keldysh line, such as $\sim\int d\omega dq G^{(R/A)}G^K$, the diffusion term ($\sim\gamma_dq^2$) mimics the two-dimensional enhancement crucial for the quantum (in $d$ dimensions) to classical (in $d+2$ dimensions, $z=2$) correspondence, while such correspondence cannot occur in diagrams with less (or more) than one $G^K$. These diagrams (cf. also Fig. \ref{oneloop}) are proportional to couplings which are RG-relevant  and non-vanishing at the quantum fixed point, and then they are  present only in the diagrammatics of $S_Q$. 



These additional  diagrams are therefore responsible for the dissimilarity between the critical exponents of the non-equilibrium quantum fixed point and of its three dimensional critical classical counterpart: in the flow of $S_Q$ there are diagrammatic corrections which do not have a counterpart in the diagrammatics of the semi-classical action, and which are beyond a naive $d$/$d+z$ correspondence.

A more pragmatical evidence is also provided by the direct comparison of the full list of critical exponents of  classical and quantum universality classes in Tab. \ref{tabella}.


\subsection{Extent of the quantum scaling regime}
\label{Sec:extent}

In this Section, we provide estimates for the lower and upper bound of the momentum window, $\Lambda_M<q<\Lambda_G$,  ($\Lambda_M$ and $\Lambda_G$ are the Markov and Ginzburg scales defined in Sec. \ref{QCanonical}), which delimits the quantum scaling regime.

\begin{figure}[t!]\centering
\includegraphics[width=6.9cm] {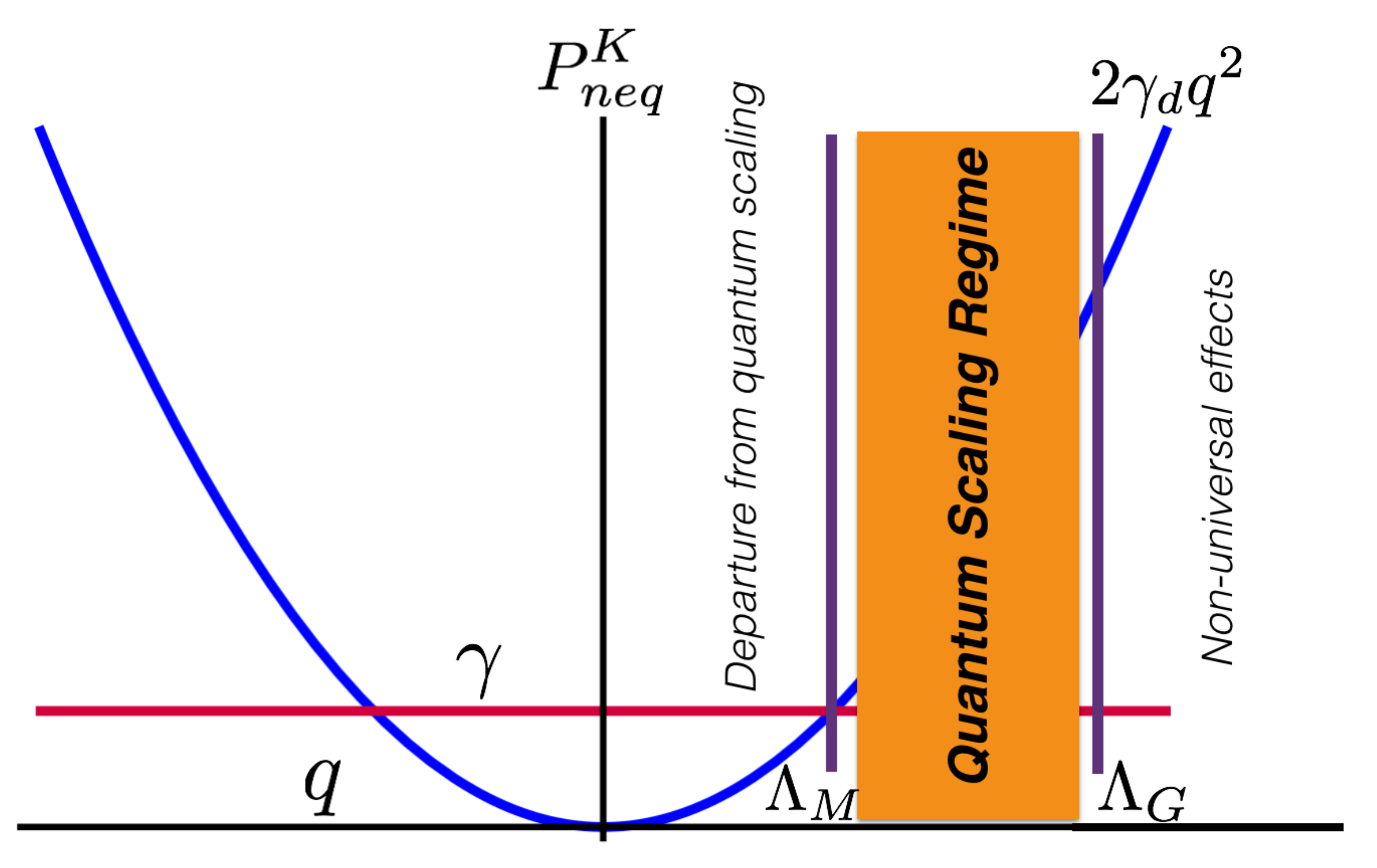}
 \caption{(Color online)  Diffusion, $\gamma_dq^2$ (blue line), and  flat, $\gamma$ (red line), Markov noises as a function of momentum. For momentum scales $q>\Lambda_G$ the RG flow is dominated by non-universal effects, while below $\Lambda_G$ it is controlled by the quantum FP, and universal effects associated to the novel non-equilibrium phase transitions emerge. For $q<\Lambda_M$, the flat Markovian noise level, $\gamma$, dominates over the  diffusion noise, and the system departs from the universal quantum scaling regime.}
\label{scales}
\end{figure}

Our estimates are based on  beta functions extracted in the symmetric phase ($\rho_0=0$ in the FRG calculations). Since the analytical extraction of $\Lambda_M$ from the beta functions in the ordered phase shows that the estimate does not substantially change the result, we decided to present the estimate in the symmetric phase, in order to make the presentation more concise. 

\subsubsection{Estimate of $\Lambda_G$}
\label{secestimG}

The beta function for the retarded mass reads in the symmetric phase,
\begin{equation}\label{betamasse}\begin{split}
\partial_t\chi&=\beta_{\chi}=-\frac{1}{\pi}\kappa_c\frac{3K_R\gamma k^3+2K_I\gamma_dk^5}{3(k^2K_I+\chi)^2}.\\
\end{split}
\end{equation}

We now integrate the flow in the momentum interval, $\Lambda_G<k<\Lambda_{UV}$ (where $\Lambda_{UV}$ is the ultra-violet momentum scale),  assuming that at micro-scales the Bose gas is supplied with a strong diffusion noise ($\gamma_dk^2\gg\gamma$), which allows us to neglect the first term on the right hand side of Eq. \eqref{betamasse}. Moreover, the solution of Eq. \eqref{betamasse} is facilitated in $\Lambda_G<k<\Lambda_{UV}$, since couplings are just perturbatively corrected, and hence we can assume that they keep their microscopic values in this momentum window (see also Tab. \ref{tabmicro})
\be\label{microginz}
\kappa_c\simeq\gamma_t,\quad g_1\simeq2\gamma_t,\quad K_I\simeq \gamma_d.
\ee 

We then expand the denominator of Eq. \eqref{betamasse} for $\chi\ll1$ (assuming that $\chi$ has been fine tuned close to zero, in order to approach the condensation transition), and we find

\begin{equation}
\partial_t\chi=-\frac{2}{3\pi}\frac{\gamma_d\kappa_c}{K_I}k\left(1-\frac{2\chi}{k^2K_I}+...\right)\simeq\frac{4}{3\pi}\frac{\gamma_d}{K_I^2}\kappa_c\frac{1}{k}\chi,
\end{equation}
where in the last equality we dropped the UV-linearly divergent term (linear term in $k$), which should not affect the flow close to $k\simeq\Lambda_G$. 
This is verified, after the value of $\Lambda_G$ has been extracted, yielding a condition for the microscopic two-body loss rate,   $\gamma_t\ll \sqrt{K_I\chi}$, with $\chi\ll1$ since the gas is fine tuned to criticality. This condition is equivalent to a perturbative assumption on the bare value of $\gamma_t$.

The resulting expression for $\chi(k)$, valid in the momentum range $\Lambda_G<k<\Lambda_{UV}$, is
\begin{equation}\label{Ginz}\begin{split}
\chi(k)&\simeq\chi_{UV}\exp\left(\frac{4}{3\pi}\frac{\gamma_d}{K_I^2}\kappa_c\left[\frac{1}{\Lambda_{UV}}-\frac{1}{k}\right)\right]\simeq \\
&\simeq_{k\ll\Lambda_{UV}}\chi_{UV}\exp\left(-\frac{4}{3\pi}\frac{\gamma_d}{K_I^2}\kappa_c\frac{1}{k}\right),
\end{split}\end{equation}
where we introduced the retarded mass at the UV scale $\chi_{UV}\equiv \chi|_{k=\Lambda_{UV}}$. 
The Ginzburg scale is found from Eq. \eqref{Ginz}
as  the  momentum scale, 
\be
k=\Lambda_G\simeq\frac{\gamma_t}{\gamma_d},
\ee
where the value of the bare (ultraviolet) retarded mass, $\chi_{UV}$, is significantly altered by loop corrections (for $k\simeq\Lambda_G$, the exponent in Eq. \eqref{Ginz} is of the order of unity, and  the value of $\chi_{UV}$ has sizeably diminished), indicating that the perturbative description breaks down. 
In order to find $\Lambda_G$, we replaced in the exponent of Eq. \eqref{Ginz}, the microscopic values of the couplings \eqref{microginz}. 


We double-checked that the same result can be found following the method provided in Ref. \onlinecite{Amit/Martin-Mayor}.

\subsubsection{Estimate of $\Lambda_M$}
\label{secestim}

In the momentum window, $\Lambda_M<k<\Lambda_G$, critical fluctuations become severe and they non-perturbatively renormalize the RG  flow of the couplings. In particular,  the Markovian noise level can receive several loop corrections, which can affect  the value of $\Lambda_M$. For this reason,  a perturbative estimate from  beta functions is inappropriate, and the estimate of $\Lambda_M$ requires the knowledge of the FP, Eq. \eqref{FPquantum}. 

This should be contrasted with finite temperature  quantum criticality, where the de-Broglie momentum scale, $\Lambda_{dB}\simeq T^{1/z}$, is given  from the outset, since temperature is invariant under RG transformations.

In the following we  provide an estimate on the upper bound of $\Lambda_M$. 

As a first task, we compute the dependence of $\gamma(k)$ on the running scale, $k$, in the vicinity of the quantum fixed point, Eq. \eqref{FPquantum}. We consider $\beta_\gamma$ in the symmetric phase, and we rescale all the couplings with their FP values, besides $\gamma(k)$ itself,
\begin{equation}\label{flowgamma}
k\partial_k \gamma=-4\tilde{g}_1^*\frac{3\gamma+2k^2\gamma_d}{3(1+\tilde{\chi}^*)^2},
\end{equation}
where we defined $\tilde{\chi}=\frac{\chi}{k^2K_I}$.
The solution of Eq. \eqref{flowgamma}, 
\be\label{gammasoluz}
\gamma(k)=-\frac{2a}{3(2+a)}\gamma_dk^2+Ck^{-a},
\ee
is parametrized in terms of the fixed point values of the rescaled couplings, in particular: $a=4\frac{{\tilde{\kappa}_c}^*{\tilde{m}^{1*}}_{\textsl{K}}}{\tilde{m}^*_R(1-\tilde{m}^*_R/4)^2}$, $C=\left(b+\tilde{\gamma}^*\right)\gamma_d\Lambda_G^{2+a}$ (with $b=\frac{2a}{3(2+a)}$). The integration constant $C$ is determined imposing the boundary condition for Eq. \eqref{flowgamma} at the Ginzburg scale,  $\gamma(k=\Lambda_G)=\tilde{\gamma}^*\gamma_d\Lambda_G^2$. 

The strategy adopted to integrate Eq. \eqref{flowgamma}, is based on the idea that for for $k\lesssim\Lambda_G$, the running couplings follow the scaling behaviour dictated by the quantum fixed point. Accordingly, the estimate for  $\gamma(k)$ cannot hold for momenta, $k\ll\Lambda_M$. 

We point out that  in a  realistic scenario, the couplings do not abruptly acquire their fixed point values  at exactly $k=\Lambda_G$ (as we implicitly assumed in the solution of Eq. \eqref{flowgamma}), but in fact this should occur at some momentum, $k\simeq p\Lambda_G$, with $p<1$.
Indeed, looking (with this more general boundary condition for Eq. \eqref{flowgamma}) for the momentum scale $k$, where the matching $\gamma(k)\simeq2\gamma_dk^2$ occurs, we find, with the help of Eq. \eqref{gammasoluz}, \begin{equation}\label{LambdaMM}
\Lambda_M=p\Lambda_G\left(\frac{{\tilde{\gamma}}^*+b}{2+b}\right)^{\frac{1}{2+a}}.
\end{equation}

Substituting the FP values of the rescaled couplings, \eqref{FPquantum}, in Eq. \eqref{LambdaMM}, we can extract the upper value estimate, $\Lambda_{M}\lesssim0.25\Lambda_G$ ($p=1$).


\section{Conclusions and outlook} 


%
In non-equilibrium classical criticality novel and richer phase diagrams of many-body driven systems (e.g. directed percolation, KPZ models, etc) are realised when the fluctuation dissipation relation is explicitly broken in the stationary state or in the dynamics\cite{tauberbook, HHRev, enz79:_field}. 
In this work, we constructed a \emph{quantum} dynamical field theory within the Keldysh formalism with the goal of analysing a non-equilibrium dynamical analog of quantum phase transitions. In particular, we considered a driven system coupled to an  environment and constrained to low dimensionality, where fluctuations and non-linear noise effects can affect non-equilibrium critical behaviour. 


Concretely, we considered a driven-dissipative Bose gas close to its critical point, and we have shown that tailoring a diffusion Markov noise, Bose condensation can be achieved in one dimension, representing a driven analog of zero-temperature quantum criticality. 

We have found the associated fixed point from  functional renormalization for driven open systems \cite{Sieberer2015}, and  extracted the  set of independent critical exponents, which characterize the universality class of this novel out-of-equilibrium critical regime. 

The salient features of the new critical regime are the survival at infrared scales of quantum  coherent effects, present already in the Lindbladian microscopic formulation of the model, and of the microscopic driven nature of the system, which breaks detailed balance. These findings constitute an important novelty in the domain of non-equilibrium criticality, since thermalization  and decoherence are expected to dominate the long wave-length physics, as reported in a broad variety of cases ranging from non-linear quantum optical to condensed matter systems \cite{Mitra2006, Diehl08, diehl10:_dynam_phase_trans_instab_open, dalla10:_quant, torre13:_keldy, Oztop2012, wouters07}. 

In particular, the absence of thermalization at the critical point, is a hallmark of the qualitative difference between the universality class discussed in this work and the one associated to the  dissipative condensation transition in three dimensions (with flat Markov noise), which constitutes its natural partner in higher dimensions. \\

We close with some perspectives. From the technical side, it is tempting to study  the $N\to\infty$ limit of a  driven-dissipative quantum action equipped with an order parameter with $N$ components ($O(N)$ models), in order to ground on the basis of exact results (the $N\to\infty$ limit is usually exactly solvable~\cite{Zinn-Justin}) the mismatch between the classical and quantum value of the correlation length critical exponent, $\nu$, for driven-dissipative criticality. Another attempt in this direction could use recent numerical advances concerning the study of the phase diagrams of driven-dissipative spin models \cite{Saro}. An accurate numerical analysis could serve to properly ground, in a variety of situations, the physical conditions for the absence of a mapping between critical quantum non-equilibrium behaviour in low dimensions and its higher dimensional classical counterparts. 
These studies would also have the potential to  scrutinise whether the absence of a mapping between quantum and classical critical behaviour out-of-equilibrium is a general feature or only a peculiar property of the fixed point analysed in this work

An other interesting direction is to inspect whether  quantum fixed points different from the one elucidated in this work, can be accessed with  non-equilibrium drivings other than the diffusion Markov noise considered here. In the presence of non-trivial conserved quantities, novel protocols to discover non-equilibrium quantum criticality can motivate further applications of the quantum dynamical field theory  and of its associated open system-FRG. However, a first necessary technical step would be to refine the truncation, Eq. \eqref{int}, introducing marginal operators, as it has been already done for the driven semi-classical condensation transition \cite{Sieberer2014}. This might be technically challenging, since  any sextic combination of the classical and quantum fields, respecting the causality structure of the Keldysh action, is allowed in this augmented truncation.\\

Finally, we also mention that the  techniques developed here might find an application in emergent quantum criticality of paired states of fermions, generated by dissipation and noise engineering \cite{Die10}. In particular, such systems can exhibit  purely noise induced critical points \cite{Eiser2010, Bar13} at finite particle density in both pure or mixed states, where the impact of  fluctuations beyond mean field is yet  unaddressed. \\



\emph{Acknowledgments.} We  thank L. M. Sieberer for support in the early stages of this project. We also acknowledge useful discussions with F. Becca, M. Buchhold, B. Capogrosso-Sansone, A. Chiocchetta, N. Dupuis, T. Gasenzer, A. Kamenev, L. He,  M. Rigol, D. Roscher and M. Vojta. J. M. acknowledges support from the Alexander Von Humboldt foundation. S. D. acknowledges funding by the German Research Foundation (DFG) through the Institutional Strategy of the University of Cologne within the German Excellence Initiative (ZUK 81), and by the European Research Council via ERC Grant Agreement n. 647434 (DOQS).


\bibliography{NEQcrit_biblio.bib}

\begin{widetext}
\newpage

\begin{appendices} 

\section{Loop corrections to quartic couplings}
\label{app:B}

In this Appendix we briefly summarise the strategy adopted to extract FRG corrections to momentum independent couplings. 
In order to make the following presentation concise, we focus on new technical aspects, referring to \onlinecite{berges02:_nonper} for general properties of functional renormalization, and to Ref. \onlinecite{Sieberer2014} for its application to critical driven open systems.\\

In order to extract corrections to momentum independent couplings,  we can manipulate   Wetterich's equation, Eq. \eqref{Wetterich}, in a form which resembles the more familiar one-loop effective potential of equilibrium field theories (see in particular Appendix C of Ref. \onlinecite{Sieberer2014}),
\begin{equation}\label{wettpot}
\frac{1}{\Omega}\partial_t\Gamma_{k,cq}=\frac{i}{2}\operatorname{Tr}\tilde{\partial}_t\log(\Gamma_k^{(2)}+R_k)=\frac{i}{2}\int_{Q}\tilde{\partial}_t \log \operatorname{det} _{cq}(\omega,q^2),
\end{equation}
where $\tilde{\partial}_t\equiv\partial_tR_{k,\bar{K}}\partial_{R_{k,\bar{K}}}+\partial_tR^*_{k,\bar{K}}\partial_{R^*_{k,\bar{K}}}$,  $\Omega$ is the quantization volume, and $\int_Q=\int\frac{d\omega d^d\mathbf{q}}{(2\pi)^{d+1}}$. With the shorthand notation  $\det_{cq}(\omega,q^2)\equiv\det (P_{cq}+R_k(q^2))$ (employed already in Ref. \onlinecite{Sieberer2014}), we labelled  the regularized inverse propagator propagator, $\Gamma_k^{(2)}$, in the presence of interactions, $\Gamma_k^{int}$ (cf. Eq. \eqref{int}). 
 
 $P_{cq}$ is a $4\times4$ matrix  for a  $O(2)$ field theory formulated in the Keldysh language, and
the subscripts ${c,q}$ indicate that $\det_{cq}(\omega,q^2)$ is a function of the homogeneous and constant background classical ($\phi_c$ and $\phi_c^*$) and quantum fields ($\phi_q$ and $\phi_q^*$). We used the same label of the classical and quantum fields for the background fields configuration, in  order to lighten notation.  

We now expand  $\log \operatorname{det} _{cq}(\omega,q^2)$ in a Taylor series around a classical stationary fixed field configuration ($\phi_c\neq0$, $\phi_c^{*}\neq0$, $\phi_q^{*}=0,\phi_q=0$, to which we refer in the following with the shorthand 'ss'),

\begin{equation}\label{taylor}\begin{split}
\log \dip_{cq}(\omega,q^2)&=a_1+a_2\phi_q+a_3\phi_q^{*}+a_4\phi_q\phi_q^{*}+\frac{a_5}{2}\phi_q^2+\frac{a_6}{2}(\phi_q^{*})^2+\\
+&\frac{1}{3!}(a_7\phi_q^3+a_8(\phi_q^{*})^3+3a_9\phi_q^2\phi_q^{*}+3a_{10}\phi_q(\phi_q^{*})^2)+\frac{a_{11}}{4!}(\phi_q\phi_q^{*})^2+... 
\end{split}
\end{equation}
We now show that the coefficients of the expansion \eqref{taylor} provide  RG corrections to  quartic couplings. \\

We start considering RG corrections to the classical coupling $g_c$, for which we require the coefficients, $a_2$ and $a_3$, of the linear terms in the quantum field  in Eq. \eqref{taylor}:
\begin{equation}
\begin{split}
a_2&=\frac{\partial}{\partial\phi_q}\ln \dip_{cq}(\omega,q^2)|_{ss}=\Big(\frac{\partial_{\phi_q}\det_{cq}(\omega,q^2)}{\det_{cq}(\omega,q^2)}\Big)\Big|_{ss}=\frac{1}{\det_c(\omega,q^2)}\partial_{\phi_q}\operatorname{det}_{cq}(\omega,q^2)|_{ss}\equiv\frac{A_q(\omega,q^2)}{\det_c(\omega,q^2)},
\end{split}
\end{equation}
\begin{equation}
a_3=\frac{1}{\det_c(\omega,q^2)}\partial_{\phi_q^{*}}\operatorname{det}_{cq}(\omega,q^2)|_{\phi_q=0,\phi_q^{*}=0}\equiv\frac{A_{qs}(\omega,q^2)}{\det_c(\omega,q^2)}.
\end{equation}
Notice that above we introduced the classical determinant $\det_c(\omega,q^2)=\det_{cq}(\omega,q^2)|_{\phi_q=0,\phi_q^{*}=0}$, which is a function of $\rho_c=\phi_c^*\phi_c$, and we labelled derivatives of $\operatorname{det} _{cq}(\omega,q^2)$ with respect to $\phi_q$ and $\phi_q^*$, with the shorthand $A_q$, $A_{qs}$, which, as a consequence of $U(1)$ invariance, can be rewritten respectively as $A_q=\phi_c^*f(\rho_c)$ and $A_{qs}=\phi_cg(\rho_c)$, where $f$, $g$ are two functions  of the  sole classical density $\rho_c$. The use of densities is a customary choice for FRG studies in the symmetric phase \cite{berges02:_nonper, Sieberer2014}. 

It is now easy to extract, for instance, the beta function of the classical coherent coupling, $\lambda_c$, projecting the left-hand side of Eq. \eqref{wettpot} onto the associated classical quartic term in Eq. \eqref{int}: this means to compute the following derivative, evaluating $\rho_c$ at its BEC-condensed value, $\rho_c=\rho_0$:
\begin{equation}\label{classicalL}
\partial_t\lambda_c= \frac{1}{2}\left(\frac{\partial^2}{\partial\rho_{cq}\partial\rho_c}+\frac{\partial^2}{\partial{\rho_{qc}}\partial\rho_c}\right)[\frac{1}{\Omega} \partial_t \Gamma_{k,cq}]|_{\rho_c\to\rho_0}= \frac{i}{2} \int_Q \dtt \frac{1}{2}\frac{\partial}{\partial{\rho_c}}\left(\frac{f(\omega,q^2,\rho_c)}{\det_c(\omega,q^2)}+\frac{g(\omega,q^2,\rho_c)}{\det_c(\omega,q^2)}\right)\Big|_{\rho_c\rightarrow\rho_0}=\Delta\lambda_c,
\end{equation}
where  we replaced the field combination $\phi_c^2\phi_c^*\phi_q^*$ (and its complex conjugate) with its density representation, $\rho_c\rho_{cq}$ ($\rho_c\rho_{qc}$), where $\rho_{cq}=\phi_c\phi_q^*$ (and $\rho_{qc}=\rho_{cq}^*$). 

With  analogue procedure  we find  for $\kappa_c$,
\begin{equation}\label{classicalK}
\partial_t\kappa_c= \frac{i}{2}\frac{\partial}{\partial{\rho_c}}\left(\frac{\partial}{\partial{\rho_{qc}}}-\frac{\partial}{\partial{\rho_{cq}}}\right)[\frac{1}{\Omega} \partial_t \Gamma_{k,cq}]|_{\rho_c\to\rho_0} = \frac{1}{2} \int_Q \dtt \frac{1}{2}\frac{\partial}{\partial{\rho_c}}\left(\frac{f(\omega,q^2,\rho_c)}{\det_c(\omega,q^2)}-\frac{g(\omega,q^2,\rho_c)}{\det_c(\omega,q^2)}\right)\Big|_{\rho_c\rightarrow\rho_0}=\Delta\kappa_c.
\end{equation}
Both in Eqs. \eqref{classicalL} and \eqref{classicalK}, the derivative wrt to the RG time, $t$, and frequency/momentum integration are taken following the standard steps outlined in Ref. \onlinecite{Sieberer2014}.

For the quantum coupling, $g_q$, we proceed in a similar fashion, and we find for instance, for the real part $\lambda_q$ (we are dropping in the following the $(\omega,q^2)$ and $\rho_c$ dependences)
\begin{equation}\label{quantproj}\begin{split}
&\partial_t\lambda_q= \frac{1}{2}\left(\frac{\partial^2}{\partial\rho_q\partial{\rho_{cq}}}+\frac{\partial^2}{\partial\rho_q\partial{\rho_{qc}}}\right)[\frac{1}{\Omega} \partial_t \Gamma_{k,cq}]\big|_{\rho_c\rightarrow\rho_0} =\\ 
&=\frac{i}{2} \int_Q  \frac{1}{2}\dtt\big(\frac{1}{\det_c^3}\frac{\rho_0}{2}fg(f+g)-\frac{1}{2\det_c^2}\frac{\rho_0}{2}(h^*f+hg)-\frac{1}{\det_c^2}(R_{(fA_{q,qs})}+\frac{\rho_c}{2}Q_{(fA_{q,qs})}+R_{(gA_{q,qs})}+\frac{\rho_c}{2}Q_{(gA_{q,qs})})+\\
&+\frac{1}{2\det_c}(R_{(l)}+\frac{\rho_c}{2}Q_{(l)}+R_{(l^*)}+\frac{\rho_c}{2}Q_{(l^*)})\big)\Big|_{\rho_c\rightarrow\rho_0}=
\Delta\lambda_q.\\
\end{split}
\end{equation}
In Eq. \eqref{quantproj} several new labels have been introduced: 
$h(\rho_c)$ is defined from $A_{q,q}A_{qs}\phi_q^2\phi^{*}_q\equiv(\phi_c^{*})^2\phi_c\phi_q^2\phi_q^{*}g(\rho_c)h(\rho_c)$ ($h^*(\rho_c)$ is its complex conjugate);
$l(\rho_c)$ is defined from $A_{q,q,qs}\phi_q^2\phi^{*}_q\equiv\phi_c^{*}\phi_q^2\phi_q^{*}l(\rho_c)$ ($l^*(\rho_c)$ is its complex conjugate);
the following concise notation for the derivatives with respect to quantum fields has been adopted, $A_{\alpha_1,\alpha_2,...,\alpha_j}\equiv{\partial^j}_{\alpha_1,\alpha_2,...,\alpha_j}\det_{cq}|_{ss}$, where $\alpha={q,qs}$, and $\partial_q$ is the short notation for $\partial_{\phi_q}$, while  $\partial_{qs}$  for $\partial_{\phi_q^*}$;
  $R_{(X(\rho_c))}$ and $Q_{(X(\rho_c))}$ stand respectively for the polynomial remainder and quotient  of $X(\rho_c)$ with respect to to the classical density $\rho_c$;
 finally, wherever a combination $\phi_q^{*}\phi_q\phi_c^{*}\phi_c$ occurs, we adopted the following symmetrization $\phi_q^{*}\phi_q\phi_c^{*}\phi_c\to\frac{1}{2}(\rho_c\rho_q+\rho_{cq}\rho_{qc})$, since $\phi_q^{*}\phi_q\phi_c^{*}\phi_c$  could be ambiguously written as $\rho_c\rho_q$ or $\rho_{cq}\rho_{qc}$.

A similar expression to Eq. \eqref{quantproj} holds for $\partial_t\kappa_q$, while in order to compute $\partial_t\gamma$ and $\partial_tg_1$ only the coefficient $a_4$ is involved. 
It is important   that the condensate density $\rho_0$ has been subtracted in Eq. \eqref{int}, since it cancels spurious contributions coming from $\partial_t{g_c}$ to $\partial_t\rho_0$, or from $\partial_tg_2$ to $\partial_t\gamma$, that would arise following the computation procedure discussed so far (see for similar aspects Ref. \onlinecite{Sieberer2014}).

 Corrections associated to $g_2$ come from the coefficients $a_{5}$ and $a_6$, which are respectively  proportional to $\phi_q^2$ and ${\phi_q^{2*}}$. Finally, we need $a_{11}$ to extract $\partial_t g_3$. All these computations follows a lengthy, yet direct, generalization of the procedure employed to compute $\lambda_q$ in Eq. \eqref{quantproj}.

The software MATHEMATICA has been employed to finalize the computations outlined in  this Appendix and find the beta functions discussed in the main text, as well as the quantum FP, \eqref{FPquantum}, and its stability matrix.

\section{Anomalous dimensions}
\label{app:C}

In this Appendix we briefly summarise the strategy adopted to extract FRG corrections to the kinetic coefficients and to $\gamma_d$, which are the momentum dependent couplings. From these corrections, we compute the anomalous dimensions introduced in Sec. \ref{flowsection}. Like for Appendix \ref{app:B}, we focus on new technical aspects, referring to \onlinecite{berges02:_nonper} and to \onlinecite{Sieberer2014} for   more conventional technical  steps.

We start with the formulas provided in Ref. \onlinecite{Sieberer2014}:
\be
\begin{split}
  \label{eq:178}
  \ezr & = \Re \eta_Z = - \frac{1}{2} \partial_{\omega} \tr ( \sigma_y \dtt \Sigma^R(Q) )
  |_{Q =(\omega,q)= 0}, \\ 
  \ezi & = \Im \eta_Z = - \frac{i}{2}
  \partial_{\omega} \tr ( \dtt \Sigma^R(Q) ) |_{Q=(\omega,q) = 0},\\
   \bar{\eta}_{K_R} & =  -
    \frac{1}{2K_R} \partial_q^2 \dtt (\Sigma^R_{22}(Q)) |_{Q=(\omega,q) = 0},\\
     \bar{\eta}_{K_I} & = - \frac{1}{2K_I} \partial_q^2 \dtt (\Sigma^R_{12}(Q)) |_{Q=(\omega,q) = 0}.
     \end{split}
\end{equation}

We find for them explicitly

\begin{equation}\begin{split}\label{Diffusione}
    \bar{\eta}_{K_R}  &= - \frac{2}{\pi K_R} k^{3} \sum_{a,b} a b \partial_a
      \partial_{b+} \sigma^R_{22+} \bigr\rvert_{q = 0,p = k}-\frac{k^{3}}{\pi K_R}\sum_a\partial_q^2|_{q=0}(\partial_a\sigma^R_{22+}|_{z_+=0,p_a=ak^2}+\partial_a\sigma^R_{22-}|_{z_-=0,p_a=ak^2})a, \\ 
      \bar{\eta}_{K_I} & =- \frac{2}{\pi K_I} k^{3} \sum_{a,b} a b \partial_a
    \partial_{b+} \sigma^R_{12+} \bigr\rvert_{q = 0,p = k}-\frac{k^{3}}{\pi K_I}\sum_a\partial_q^2|_{q=0}(\partial_a\sigma^R_{12+}|_{z_+=0,p_a=ak^2}+\partial_a\sigma^R_{12-}|_{z_-=0,p_a=ak^2})a,
  \end{split}
\end{equation}
where $a,b=K_R,K_I$, the abbreviation $\partial_a$ stands for $\partial_{p_a(x)}$ and $\partial_{a\pm}\equiv\partial_{p_a(x_{\pm})}$, $x_\pm\equiv|p\pm q|^2$, $z_{\pm}\equiv(x_{\pm}-k^2)\theta(x_{\pm}-k^2)$ and $\sigma_{i,j}$ indicates the frequency-integrated self-energy, $\int\frac{d\omega}{2\pi}\Sigma_{i,j}$ (recall that $\Sigma$ is in the Keldysh basis a $4\times4$ matrix, whose entries depend on frequency and momentum; for the explicit expression see  Eq. \eqref{self} below). This nomenclature has been chosen consistently with previous FRG literature \cite{wetterich08:_funct, Sieberer2014}.

Because of the additional momentum dependence coming  from quantum diffusion, we find an additional correction (second term in the RHS of Eq. \eqref{Diffusione}) to the results provided in \cite{Sieberer2014}. In particular, $\partial^2_q$ suppresses $\gamma_d$-independent terms, and selects only contributions coming from noise diffusion.

For the anomalous dimension of $\bar{\gamma}_d$ we get $\bar{\eta}_{\gamma_d}=\frac{1}{8i\gamma_d}\partial_{q^2}\dtt\left(\Sigma^K_{11}+\Sigma^K_{22}\right)|_{q=0}$, which can be manipulated, like $\bar{\eta}_{\{KR,KI\}}$, following  Appendix C of Ref. \onlinecite{Sieberer2014}, resulting in  expressions analogous to Eqs. \eqref{Diffusione}.


We now report  the expressions of the self-energies involved in the computations of the anomalous dimensions  (we define $P_+=P+Q$, where $P=(p,\omega')$)
\begin{equation}\label{self}
\begin{split}
  &\Sigma^R_{22}(Q) = \frac{i}{2} \int_P \tr \left( G^K(P)v_4^{I}G^R(P_+)v_2^{II} + G^A(P)v_4^{I}G^K(P_{+})v_2^{II} \right),\\
  &\Sigma^R_{12}(Q) = \frac{i}{2} \int_P \tr \left( G^K(P)v_3^{I}G^R(P_+)v_2^{II} + G^A(P)v_3^{I}G^K(P_{+})v_2^{II} \right),\\
 &\Sigma^K_{11}(Q) = \frac{i}{2} \int_P \tr \big(  G^K(P)v_3^{I}G^K(P_+)v_3^{I} + G^K(P)v_3^{I}G^R(P_+)v_3^{II} +
 G^R(P)v_3^{II}G^K(P_+)v_3^{I} + G^K(P)v_3^{II}G^A(P_+)v_3^{I} +\\
 &+G^R(P)v_3^{III}G^A(P_+)v_3^{I} + G^A(P)v_3^{I}G^K(P_+)v_3^{II} +
 G^A(P)v_3^{I}G^R(P_+)v_3^{III} \big),\\
& \Sigma^K_{22}(Q) = \frac{i}{2} \int_P \tr \big( G^K(P)v_4^{I}G^K(P_+)v_4^{I} + G^K(P)v_4^{I}G^R(P_+)v_4^{III} +
 G^R(P)v_4^{III}G^K(P_+)v_4^{I} + G^K(P)v_4^{II}G^A(P_+)v_4^{I} +\\
 &+G^R(P)v_4^{IV}G^A(P_+)v_4^{I} + G^A(P)v_4^{I}G^K(P_+)v_4^{II} +
 G^A(P)v_4^{I}G^R(P_+)v_4^{IV} \big), 
\end{split}
\end{equation}

where for the interaction vertices we have 
\begin{equation}\begin{split}
v_{1}^{II}&=\begin{pmatrix} 3\sqrt{2}\lambda_c\sqrt{\rho_0} & -3\sqrt{2}\kappa_c\sqrt{\rho_0}\\
\sqrt{2}\kappa_c\sqrt{\rho_0} & \sqrt{2}\lambda_c\sqrt{\rho_0}\end{pmatrix},\quad
v_{1}^{III}=\begin{pmatrix} 3\sqrt{2}\lambda_c\sqrt{\rho_0} & \sqrt{2}\kappa_c\sqrt{\rho_0}\\
-3\sqrt{2}\kappa_c\sqrt{\rho_0} & \sqrt{2}\lambda_c\sqrt{\rho_0}\end{pmatrix},\quad
v_{1}^{IV}=\begin{pmatrix} -4i\sqrt{2}\sqrt{\rho_0}g_1 & 0\\
0 & -4i\sqrt{2}\sqrt{\rho_0}g_1 \end{pmatrix},\\
v_{2}^{II}&=\begin{pmatrix} \sqrt{2}\kappa_c\sqrt{\rho_0} &\sqrt{2}\lambda_c\sqrt{\rho_0}\\
\sqrt{2}\lambda_c\sqrt{\rho_0} & -\sqrt{2}\kappa_c\sqrt{\rho_0}\end{pmatrix},\quad
v_{2}^{IV}=v_{2}^{II},\\
v_{3}^{I}&=\begin{pmatrix} 3\sqrt{2}\lambda_c\sqrt{\rho_0} & \sqrt{2}\kappa_c\sqrt{\rho_0}\\
\sqrt{2}\kappa_c\sqrt{\rho_0} & \sqrt{2}\lambda_c\sqrt{\rho_0}\end{pmatrix},\quad
v_{3}^{II}=\begin{pmatrix} -4i\sqrt{2}g_1\sqrt{\rho_0} &0\\
0 & 0\end{pmatrix}=v_{3}^{III},\quad
v_{3}^{IV}=\begin{pmatrix} 3\sqrt{2}\lambda_q\sqrt{\rho_0}&\sqrt{2}\kappa_c\sqrt{\rho_0}\\
\sqrt{2}\kappa_q\sqrt{\rho_0}&\sqrt{2}\lambda_q\sqrt{\rho_0}\end{pmatrix},\\
v_{4}^{I}&=\begin{pmatrix} -3\sqrt{2}\kappa_c\sqrt{\rho_0} & \sqrt{2}\lambda_c\sqrt{\rho_0}\\
\sqrt{2}\lambda_c\sqrt{\rho_0} & -\sqrt{2}\kappa_c\sqrt{\rho_0}\end{pmatrix},\quad
v_{4}^{II}=\begin{pmatrix}0& -4i\sqrt{2}g_1\sqrt{\rho_0}\\
0 &0\end{pmatrix},\quad
v_{4}^{III}=\begin{pmatrix} 0 &0\\
-4i\sqrt{2}g_1\sqrt{\rho_0} & 0\end{pmatrix},\quad\\
v_{1}^{IV}&=\begin{pmatrix} \sqrt{2}\sqrt{\rho_0}\kappa_q & \sqrt{2}\sqrt{\rho_0}\lambda_q\\
\sqrt{2}\sqrt{\rho_0}\lambda_q & 3\sqrt{2}\sqrt{\rho_0}\kappa_q \end{pmatrix}.
\end{split}
\end{equation}

\end{appendices}

\end{widetext}

\end{document}